\documentclass[manuscript,nonacm]{acmart}

\usepackage[dvipsnames]{xcolor}
\usepackage{multirow}
\usepackage{subcaption}
\AtBeginDocument{%
  }

\setcopyright{none}
\renewcommand{\footnotetextcopyrightpermission}[1]{}

\acmConference[Preprint]{UK}{September}{2026}
\begin{document}

\title{When AI Becomes Hard to Understand: Cognitive Demands in Real-World Human–AI Conversations}


\author{Yingcan Carol Wang}
\email{carol.wang@stripepartners.com}
\affiliation{%
  \institution{Stripe Partners}
  \city{}
  \country{United Kingdom}
}

\author{Iman Munire Bilal}
\email{iman.bilal@stripepartners.com}
\affiliation{%
  \institution{Stripe Partners}
   \city{}
  \country{United Kingdom}
}

\author{Qamar Zaman}
\email{qamar.zaman@stripepartners.com}
\affiliation{%
  \institution{Stripe Partners}
  \city{}
  \country{United Kingdom}
}

\renewcommand{\shortauthors}{Wang et al.}

\begin{abstract}
 Generative AI increasingly supports complex financial and health decisions, yet we know little about when its responses become difficult to process in real-world dialogue. We analyse more than 84,000 ChatGPT and Gemini conversations, using repeated prompting and clarification following misunderstanding as behavioural indicators of cognitive difficulty. We find that response characteristics such as length, readability and lexical diversity do not have fixed relationships with conversational difficulty; instead, their relationships depend on how they combine. Most notably, greater lexical diversity was associated with less repeated prompting in shorter responses, but this association weakened as response length increased, a pattern that replicated across financial and health conversations. We propose a conversational complexity budget to conceptualise these interdependencies: the demands associated with one response characteristic may depend on those accompanying it. The resulting design challenge is how to configure response complexity for the particular user, task and interaction.
\end{abstract}






\maketitle

\section{Introduction}

Generative AI systems increasingly mediate tasks that require users to understand unfamiliar concepts, compare alternatives, and interpret complex evidence \cite{chen2025more}. This responsibility is acute in domains such as Health and Finance, where users may act on AI-generated information \cite{our_paper1, costagomes2026health}. When outputs become hard to process, the consequences may extend beyond interaction friction: misunderstanding or incomplete comprehension can undermine the judgements and decisions that AI is being used to support \cite{bucinca2021, friedman2026shortlongllmresponse}. 

Conversational AI makes this problem particularly challenging because the information environment is dynamically generated. Responses vary in their length, language, information content, and structure, while users approach the same interface with tasks requiring very different forms of cognitive work \cite{wu2026textual,collins2014,McNamara2012,Parasuraman2000}. We know comparatively little about how these sources of cognitive demand relate to difficulty in naturally occurring AI interactions: how difficulty varies across user goals, and whether response characteristics operate independently or in combination. Studying these questions at scale is difficult because established approaches to cognitive load typically rely on controlled tasks, self-report, or physiological measurement \cite{cruz2026temporal,lepine2026precision,hart1988development}. Conversational interaction, however, leaves behavioural traces when understanding breaks down: users may repeat or reformulate a request, or explicitly ask the system for clarification \cite{ALLOATTI2024101603,DIPPOLD202321,FREEMAN201812}. Such repair behaviours do not directly measure cognitive load, but provide observable indicators of moments where conversational difficulty becomes severe enough to disrupt the interaction.

In this paper, we investigate these questions using more than 84,000 naturally occurring generative-AI conversations across Finance and Health. We identify two forms of conversational repair, fragmented repeated prompting and clarification following apparent misunderstanding, and use them as behavioural indicators of conversational difficulty. We examine how these behaviours vary with users’ interaction goals and substantive context, and with linguistic and structural characteristics of AI responses, including response length, readability, lexical diversity, information density, and formatting. We then model interactions between response characteristics to examine whether their relationships with conversational breakdown depend on how responses are configured.

Across both domains, conversational breakdown differed across user intents but did not follow a simple gradient with the level of decision authority delegated to AI. Higher-order research, comparison, and personalised-analysis intents were associated with elevated breakdown, while simple retrieval was associated with lower breakdown. Yet learning was also associated with elevated breakdown despite involving relatively little delegation of judgement. 

Individual response characteristics also did not map straightforwardly onto difficulty: longer or more linguistically diverse responses were not uniformly more difficult, while greater visual structuring was not uniformly protective. Instead, their relationships with difficulty sometimes depended on other response characteristics. Most notably, across both Finance and Health, greater lexical diversity was associated with less repeated prompting when responses were relatively short, but this advantage diminished as responses became longer. These results suggest that conversational difficulty is not adequately characterised by simple rules about individual response features, but varies with the cognitive work implied by users’ goals and with how demands within the response combine.

This paper makes three contributions to HCI research on cognitive demands in human–AI interaction. 

\begin{itemize}
    \item Methodologically, we show how conversational repair behaviours can identify critical friction points across large-scale, naturally occurring AI interactions, complementing experimental and self-report measures of cognitive load.
    \item Empirically, we show that conversational difficulty is task-dependent and configurational: rather than mapping simply onto individual response features such as length or readability, it is associated with how particular response characteristics combine. 
    \item Conceptually, we propose a conversational complexity budget for understanding how multiple response demands may combine, and argue that comprehensibility is better understood in relation to the user and task than as a property of the response alone. 
\end{itemize}

%[include research questions in this section]

\section{Related Work}
\label{sec:related_work}

\subsection{Cognitive load in human–AI interaction}

Cognitive Load Theory conceptualises human information processing as constrained by limited working-memory capacity ~\cite{sweller1998cognitive,Sweller2019}. The demands placed on this capacity stem not just from information volume, but from element interactivity, the number and interdependence of elements processed simultaneously, which means that similar amounts of information can impose very different processing demands \cite{leahy2019centrality,Sweller2010}.

CLT further distinguishes between demands arising from the inherent complexity of the material and avoidable demands introduced by how information is structured or presented. The latter are particularly relevant to interface design: poorly organised or unnecessarily difficult presentation can consume cognitive resources without advancing the user's underlying task. The effective complexity of information also depends on what the individual already knows. Research on the expertise-reversal effect, for example, shows that forms of instruction helpful to novices may become redundant or counterproductive for more knowledgeable users \cite{kalyuga2003expertise}.

These principles take on a particular form in conversational interfaces, where information is generated dynamically over the course of an interaction. Unlike fixed information displays, AI conversations unfold sequentially: users must process generated information while maintaining the context of the interaction and determining what to ask or do next. Prior work integrating CLT with HCI highlights the importance of designing information environments around both task demands and users' limited processing resources \cite{hollender2010integrating}. Generative AI raises this problem in a particularly open-ended form, because response length, linguistic complexity, information density and structure can vary substantially from one interaction to another.

In this study, we use Cognitive Load Theory as a theoretical lens for understanding how task and response characteristics may generate cognitive processing demands. We do not directly measure cognitive load or attempt to decompose it into intrinsic and extraneous components.

\subsection{Conversational breakdown and repair as behavioural evidence of difficulty}

Measuring cognitive difficulty in naturalistic human–computer interaction is challenging. Established measures such as NASA-TLX provide useful subjective assessments but are typically collected after task performance, while physiological approaches such as eye-tracking or neurophysiological measurement require instrumentation that is difficult to deploy at scale \cite{hart1988development,lepine2026precision}. These constraints are particularly relevant when studying large-scale, naturally occurring interactions outside controlled settings.

Recent HCI research has therefore begun to infer cognitive and metacognitive states from interaction traces themselves. Cruz et al. \cite{cruz2026temporal}, for example, used observable features of human–AI conversations to infer latent cognitive states, while Lepine et al. \cite{lepine2026precision} developed transcript-based computational indicators of intrinsic and extraneous cognitive load based on interaction characteristics such as task switching, information sprawl, and dependency debt. This work demonstrates the potential for conversational traces to provide scalable evidence about cognitive processes without requiring retrospective self-report or physiological instrumentation.

Conversation analysis defines repair as the organised practices through which interlocutors address problems in speaking, hearing, or understanding \cite{Schegloffetal1977}. Repair behaviours are also observed in human–AI interaction. When conversational agents fail to understand or misinterpret users, people respond through strategies including repetition, rephrasing, and clarification \cite{DIPPOLD202321}. Recent work has further shown that such repair strategies can be identified computationally from conversational-agent traces \cite{ALLOATTI2024101603}.

User restatement is particularly relevant to our study. Freeman and Beaver \cite{FREEMAN201812} examined restatements in nearly 3,000 interactions with a deployed virtual assistant and found that users may restate a request even when there is no apparent system misunderstanding. Their results suggest that restatement can reflect problems with how a response is formulated or presented, making it potentially informative about whether a response enables the interaction to progress successfully.

We examine two repair behaviours observable directly in conversational traces: fragmented repeated prompting, in which users substantially repeat or reformulate an earlier request, and clarification following apparent misunderstanding, in which users explicitly seek clarification of an AI response. We treat these behaviours as scalable indicators of conversational difficulty rather than direct measures of cognitive load. Both identify moments when difficulty becomes sufficiently pronounced to disrupt the interaction, while remaining observable in large-scale, naturally occurring conversational data.

\subsection{AI response characteristics and cognitive demand}

The design and presentation of information can itself impose processing demands, making the characteristics of generated responses a potential source of conversational difficulty. Existing research has linked properties such as response length and information volume to cognitive load and continued system use. Wu et al., for example, find experimentally that longer dialogue and higher information volume increase cognitive load and, in turn, users’ intentions to discontinue conversational AI \cite{wu2026textual}. More broadly, research on text processing and readability shows that processing difficulty depends on multiple linguistic and discourse characteristics rather than text length alone \cite{collins2014,McNamara2012}.

The relationship between response characteristics and successful processing may not be monotonic. Friedman et al. experimentally varied the length of LLM-generated explanations and found that, when the AI’s reasoning was incorrect, medium-length explanations supported more accurate error detection than either shorter or longer explanations \cite{friedman2026shortlongllmresponse}. Response complexity is also multidimensional: length captures the quantity of material presented; readability captures aspects of linguistic difficulty; lexical diversity reflects variation in vocabulary; information density captures how much substantive content is compressed into the text; and formatting alters how information is visually organised. These dimensions need not operate independently.

Yet existing work has typically examined individual presentation characteristics or experimentally specified response formats, leaving comparatively little evidence about how multiple dimensions of generated responses operate in combination during naturally occurring human–AI conversation. We therefore examine whether the relationship between a given response characteristic and conversational difficulty depends on other characteristics of the response.

\subsection{Task demands and cognitive engagement in human–AI interaction}

Cognitive demands also depend on what users are trying to accomplish. Human–AI interactions encompass diverse task types, including retrieving bounded information, learning unfamiliar material, comparing alternatives, undertaking complex analysis, and delegating elements of judgement or action. These activities impose different forms of cognitive work on the user even when they occur within the same interface.

Prior work on human–automation interaction has long conceptualised automation in terms of how cognitive and operational functions are allocated between human and system, from information acquisition and analysis through decision selection and action implementation \cite{Parasuraman2000}. More recent research on human–AI interaction demonstrates that increasing AI assistance does not necessarily translate straightforwardly into better cognitive outcomes. Chen et al., for example, show that highly automated AI-supported note-taking can reduce immediate effort while also reducing subsequent comprehension and retention, whereas intermediate forms of assistance can preserve greater cognitive engagement \cite{chen2025more}. Related HCI work similarly shows that the presentation of AI assistance can alter users' cognitive engagement and reliance \cite{bucinca2021}.

This work points to a distinction between the decision authority delegated to AI and the processing demands placed on the user. These dimensions need not move together: an interaction involving little delegation may still require substantial user processing, for example when the goal is to understand unfamiliar information.

\subsection{Research gap}
Prior work establishes that cognitive demand depends on both the information being processed and the context in which it is processed; that conversational difficulty can become visible through repair behaviours; and that AI interface and response design can shape cognitive engagement. 

We know comparatively little about cognitive difficulty in naturally occurring generative-AI conversations at scale, where users pursue heterogeneous goals rather than completing predefined experimental tasks. Existing work also tends to examine response characteristics such as length, readability, and information density individually, leaving unclear how multiple dimensions of generated responses operate in combination. Finally, the growing ability of AI to move from providing information toward supporting and undertaking judgement raises an unresolved question about the relationship between decision authority delegated to AI and cognitive work demanded from the user.

We address these gaps by analysing naturally occurring Finance and Health conversations and examining how observable conversational breakdown relates to user intent, substantive domain, and the linguistic and structural characteristics of AI responses.

\section{Data overview and processing}
\subsection{Data source}

We extend the dataset introduced by Bilal et al. \cite{our_paper1}, which contains naturalistic ChatGPT and Gemini conversation histories collected from US-based users recruited and reimbursed through Measure Protocol on a voluntary opt-in basis. The corpus records user messages, AI responses, and associated timestamps. We extend the observation period to July–October 2025, yielding data from 2,626 users and more than 1.1 million user prompts.

% The dataset of Human-AI conversations introduced by \cite{our_paper1} is used as the basis of our experiments. This covers ChatGPT and Gemini conversations from 2499 US users spanning the period August-October 2025 comprising more than 1.5 million user prompts.

\subsection{Conversation segmentation and processing}

Raw conversation histories do not necessarily correspond to single coherent interactions: users may pursue multiple unrelated topics within the same AI session. Following \cite{our_paper1}, we therefore segmented longer conversation histories into topically coherent units, termed subchats. The segmentation procedure uses lexical overlap between consecutive user prompts to identify topic continuity, while accounting for short prompts that contain insufficient information to determine a topic change. Sessions containing ten or fewer user prompts were retained as single subchats. In longer sessions, consecutive prompts sharing low-frequency terms were retained within the same subchat, while longer prompts without such overlap initiated a new subchat. This procedure yielded approximately 395,000 subchats, which form the conversation-level units used in the subsequent analysis. We refer to these units as conversations hereafter.

Each turn was identified by its authoring entity (<user> or <assistant>), allowing user prompts and AI responses to be processed separately. As part of data cleaning, we removed empty turns and stripped leading and trailing whitespace. We otherwise retained the original text for construction of the linguistic measures described in Sections \ref{section:user_measures} and \ref{section:ai_measures}.

%The unit of analysis was a subchat comprising an ordered sequence of user and AI-assistant turns. We analyzed labelled finance and health conversations. Figure/Table [X] reports the number of subchats, users, user turns and assistant turns in each dataset respectively.

% Transcripts contained explicit `<user>` and `<assistant>` role markers. We recovered assistant turns by extracting the text following each `<assistant>` marker and preceding the next `<user>` marker. User turns were extracted analogously from the text following each `<user>` marker and preceding the next `<assistant>` marker. We removed empty turns and stripped leading and trailing whitespace. We did not otherwise alter the text before calculating readability, response length, formatting, lexical diversity, or information-density measures.

% The pipeline also identified the first and last substantive assistant responses for descriptive purposes. If the first extracted assistant response began with the string “search,” it was treated as a system-generated search trace and excluded from this first/last-response representation. These fields were not used to calculate the primary event-bounded linguistic measures.

\subsection{Domain selection and identification}

Instead of considering the full corpus of Human–AI conversations, we focus on Finance and Health. Both are high-stakes domains in which LLMs are increasingly used to support information seeking and decision-making, as evidenced by prior analyses of Microsoft Copilot interaction logs \cite{costagomes2025itstimetemporalmodal, costagomes2026health} and ChatGPT and Gemini conversations \cite{our_paper1}. Human–AI interactions in these domains may also involve substantial cognitive demands, raising the possibility that cognitive load could impair users’ ability to understand, evaluate, and act on AI-generated information, with potentially important consequences for their financial and health outcomes.

We identified Finance and Health conversations by adapting the domain-classification approach of \cite{our_paper1}. We constructed a manually annotated training dataset by sampling 4000 subchats from the full corpus. Two members of the research team labelled each sampled subchat as Finance, Health, or Other. We then fine-tuned a Longformer classifier \cite{Beltagy2020Longformer} on the dataset using a multi-class setting (4 epochs, learning rate = 2e-5, batch size = 4) and applied the trained model to the full corpus.

The resulting analytical datasets \footnote{For both Health and Finance, we exclude  conversations with no user-prompt word before the cognitive load event. For Health, we additionally require each conversation to contain at least one relevant content-taxonomy label and at least one behavioural-intent label (see Section 4.1).} contain approximately 43,100 Finance conversations from 1,745 users and 41,500 Health conversations from 1,558 users (see Table \ref{tab:corpus_stats}). These domain-specific datasets form the basis of the analyses that follow.

\begin{table}[]
    \centering
    \begin{tabular}{lcc}
    \toprule
    Size of domain & Finance & Health \\
    \midrule
    \# of conversations & 43.1k & 41.5k \\
    %\% of conversations out of full dataset & & \\
    \# of users &1745 & 1558 \\
    %\% of users out of full user sample &
    \midrule
    \# of ChatGPT conversations &41.3k & 40.4k \\
    \# of Gemini conversations & 1813 & 1185 \\

    \bottomrule
    \end{tabular}
    \caption{Size of Health and Finance datasets in Human-AI conversations from July-October 2025}
    \label{tab:corpus_stats}
\end{table}

\subsection{Ethics and Privacy}

This study involves secondary analysis of human–AI conversations voluntarily contributed through Measure Protocol, a third-party behavioural-data provider. Participants explicitly opted in to data-sharing tasks, chose whether to contribute their ChatGPT or Gemini interaction histories for research purposes, and were compensated for doing so. Data were not obtained through passive scraping or background monitoring.

The researchers received de-identified conversation data and did not have access to participants’ direct identifiers or attempt to re-identify individuals. Given the potentially sensitive nature of Finance and Health conversations, access to the data was restricted to authorised research team members within secure, access-controlled systems. Raw conversational data were used only where required for computational processing, annotation and validation; subsequent analyses primarily used derived classifications and linguistic measures. Results are reported only in aggregate.

\section{Methodology}
Our methodology comprises four components: classifying conversations by substantive content and user intent; constructing two conversational repair measures as behavioural indicators of cognitive difficulty; characterising the linguistic and structural properties of AI responses and their potential processing demands; and modelling how response characteristics, user intent, and substantive context are associated with these repair behaviours.

%\subsection{Data Processing}

%\subsubsection{Chat Segmentation}
%[Iman to add...]

\subsection{Content and Intent Classification}
We characterise conversations along two complementary dimensions. Content categories capture what the conversation is about, while intent categories capture what the user is attempting to accomplish with the AI. Both are non-mutually exclusive: a conversation can concern multiple substantive categories and can contain more than one user intent.

\subsubsection{Finance and Health categories}

\emph{Finance category labelling}: For Finance, we use the hierarchical taxonomy and labelling procedure developed by \cite{our_paper1}, comprising nine broad categories and 27 subcategories covering areas such as Investments, Retail Banking \& Credit, Insurance, Payments \& Transfers, Tax and Business Finance (see Appendix \ref{appen:finance_taxonomy} for more details).

\emph{Health category labelling}: For Health, we developed an analogous hierarchical, multi-label taxonomy comprising both general health content and disease-specific content. The general-health taxonomy contains five broad categories and 18 subcategories informed by prior health typologies \cite{SMAILHODZIC2021114469,scepanovic2021healthy,rivas2020}. The disease taxonomy contains seven broad categories and 25 subcategories based on the World Health Organization's International Classification of Diseases \cite{harrison2021icd11} (see Appendix \ref{appen:health_taxonomy} for more details). Each taxonomy entry linked one or more lexical variants to a canonical health entity, subcategory, and broad category.

%We case-normalized each conversation, replaced non-alphanumeric characters with spaces, and collapsed repeated whitespace. We then matched contiguous token sequences against a predefined dictionary containing 364 lexical variants representing 204 canonical entities for health and 2.7k lexical variants representing 2.3 canonical entities for finance. Matching respected token boundaries and prioritized longer multiword expressions over shorter expressions. For example, “high blood pressure” was normalized to the canonical entity hypertension and assigned to Cardiovascular and Circulatory Diseases within the broader Non-Communicable Systemic and Organ-Specific Diseases category. The procedure labeled only entities explicitly mentioned in the conversation and did not infer unobserved diagnoses or health states.

We case-normalised conversation text, replaced non-alphanumeric characters with spaces, and collapsed repeated whitespace. Contiguous token sequences were then matched against predefined dictionaries\footnote{Health dictionary contains 364 lexical variants representing 204 canonical entities. Finance dictionary contains 2.7k lexical variants representing 2.3k canonical entities}. Matching respected token boundaries and prioritised longer multiword expressions over shorter expressions. The procedure labels only entities explicitly present in the conversation; it does not infer unobserved diagnoses or health states. 

Labels were non-mutually exclusive, so a conversation may contain multiple entities and receive multiple category or subcategory labels. Repeated detections of the same canonical entity within a taxonomy location were deduplicated. For each conversation, we retained the canonical entity names, general categories and subcategories, disease categories and subcategories, and the number of unique hierarchical entity records detected.

\subsubsection{Intent categories}
We define intent as the topic-independent task a user is attempting to accomplish with the LLM. We build on the behavioural-intent framework introduced by Bilal et al. \cite{our_paper1}, which distinguishes user goals such as simple information retrieval, learning, comparison, analysis, planning, problem resolution and execution. The framework also maps intents onto three levels of decision authority, representing the extent to which decision-making responsibility is delegated to the AI: Inform, where AI primarily provides information or explanation; Shape, where AI supports analysis, judgement or choice; and Act, where AI is asked to execute or operationalise an action.

We retain the Financial Services intent taxonomy from \cite{our_paper1} and adapt it for the Health domain. 

Because execution-related intents were rare in the observed conversations, we combine Delegated Execution, Instruction-led Execution, and Automation \& Monitoring into a single Execution \& Automation category. The complete domain-specific intent taxonomies are reported in Appendix \ref{appen:intent_taxonomy}.

To capture the semantic diversity associated with each intent, we fine-tuned BigBird \cite{bigbird}, a Transformer model capable of processing long-form text, for intent classification. We formulated this as a multi-label task because a single conversation can contain more than one user goal. We trained separate intent classifiers for Finance and Health because the two domains contain substantially different substantive contexts. For each domain, we sampled conversations from the corresponding corpus to construct a training dataset and augmented these with synthetic conversations generated by GPT-5.5 to increase representation of low-prevalence intents, following \cite{our_paper1}. The final training dataset comprised 6,184 conversations spanning all intent types. Four members of the research team annotated the training data using the intent taxonomy, with a subset labelled jointly to refine and align the annotation guidelines.

Once the training datasets were finalised, we fine-tuned BigBird on 90\% of the labelled data for eight epochs in Finance and five epochs in Health. The remaining 10\% was used for evaluation and threshold tuning based on macro-averaged F1. The best Finance checkpoint achieved a $F1_{Macro}$ of 0.723 and the Health checkpoint 0.701. We then applied the fine-tuned classifiers to the full Finance and Health datasets to assign conversation-level intent labels.

% \subsection{Analysis Unit}
% The unit of analysis was a subchat comprising an ordered sequence of user and AI-assistant turns. We analyzed labelled finance and health conversations. Figure/Table [X] reports the number of subchats, users, user turns and assistant turns in each dataset respectively.

% Transcripts contained explicit `<user>` and `<assistant>` role markers. We recovered assistant turns by extracting the text following each `<assistant>` marker and preceding the next `<user>` marker. User turns were extracted analogously from the text following each `<user>` marker and preceding the next `<assistant>` marker. We removed empty turns and stripped leading and trailing whitespace. We did not otherwise alter the text before calculating readability, response length, formatting, lexical diversity, or information-density measures.

% The pipeline also identified the first and last substantive assistant responses for descriptive purposes. If the first extracted assistant response began with the string “search,” it was treated as a system-generated search trace and excluded from this first/last-response representation. These fields were not used to calculate the primary event-bounded linguistic measures.

\subsection{Behavioural Indicators of Cognitive Difficulty}
\label{section:user_measures}
We developed two conversational repair measures as behavioural indicators of cognitive difficulty: fragmented repeated prompting and clarification following AI-induced misunderstanding. The measures capture different manifestations of difficulty. Repeated prompting is a relatively broad indicator that a conversation has failed to progress as intended, whereas clarification following AI-induced misunderstanding more specifically identifies instances in which users explicitly signal difficulty understanding or making sense of an AI response.

\subsubsection{Fragmented Repeated Prompts}
We developed a rule-based procedure to identify instances in which a user repeated or reformulated an earlier prompt within the same conversation. Such repetition can indicate that the preceding interaction did not adequately resolve the user's request, prompting the user to restate or reformulate it. Because lexical similarity can also arise from short acknowledgements and conversational backchannels, the procedure was designed to exclude these cases.

User prompts were lowercased, and URLs were removed before tokenization. A prompt was eligible for repetition detection when it:
\begin{itemize}
    \item contained at least two content tokens after normalization;
    \item contained at least 80\% ASCII characters among its alphabetic characters;
    \item was not primarily an acknowledgement, greeting, thanks, agreement, disagreement, or other short interactional response; and
    \item did not contain one of the predetermined excluded phrases used for feature notifications or simple confirmations.  
\end{itemize}

%The exclusion dictionary included common forms such as “yes,” “okay,” “thanks,” and related variants. Transcript annotations such as “prompted,” “prompt,” and “unprompted” were also removed.

Eligible prompts were normalized through four steps. First, URLs, punctuation, and non-alphanumeric characters were removed. Second, common English prompt stopwords were excluded. Third, a lightweight rule-based stemmer normalized selected plural and inflectional endings. %, including `-ies`, `-ing`, `-ed`, and plural `-s`. 
Fourth, health-domain terms were mapped to canonical concepts using a predefined synonym dictionary constructed from the study’s health and disease taxonomy. Taxonomy terms not included in a synonym group retained their normalized form.

%The complete stopword list, exclusion dictionary, health taxonomy, and synonym mapping are provided in the archived analysis code and \colorbox{yellow}{[Supplementary Table/File X]}.

For every eligible prompt \(p_i\), we compared its normalized representation with all preceding eligible prompts \(p_j\) in the same conversation. We calculated: character-sequence similarity between the normalized token strings using Python’s 'SequenceMatcher'; Jaccard similarity between their unique-token sets; and token containment relative to the shorter prompt.

For token sets \(T_i\) and \(T_j\), Jaccard similarity was:

\[
J(T_i,T_j)
=
\frac{|T_i\cap T_j|}
{|T_i\cup T_j|},
\]

and containment was:

\[
C(T_i,T_j)
=
\frac{|T_i\cap T_j|}
{\min(|T_i|,|T_j|)}.
\]

Pairs sharing fewer than two unique content tokens received a similarity score of zero. Otherwise, the composite similarity was:

\[
S(p_i,p_j)
=
\max\left(
S_{\mathrm{sequence}},
J(T_i,T_j),
0.85C(T_i,T_j)
\right).
\]

A prompt was classified as a repetition if its maximum similarity to any preceding eligible prompt was at least 0.72. This threshold was derived empirically based on manual evaluation of candidate repetitive prompts in a prior pilot study.

\[
R_i
=
\mathbb{1}
\left[
\max_{j<i}S(p_i,p_j)\geq0.72
\right].
\]

The comparison with all preceding prompts allowed the method to detect a return to an earlier request even when unrelated prompts occurred between the original and repeated request.

For each conversation, we calculated: the total number of user prompts, the number of prompts classified as repetitions, the proportion of prompts classified as repetitions and the longest run of consecutively similar prompts. In addition, we extracted  up to three example prompt pairs for audit and error analysis in pilot studies.

The repetition proportion was:

\[
P_{\mathrm{repeat}}
=
\frac{\sum_i R_i}
{N_{\mathrm{user\ prompts}}}.
\]

For non-empty conversations, the denominator included all user prompts, including prompts ineligible for pairwise repetition classification. Conversations without user prompts used a computational denominator of one and consequently received a repetition proportion of zero.

Consecutive repetition runs were computed separately by comparing each prompt only with its immediately preceding prompt. This run-length measure was supplementary; the primary repetition count compared each prompt with all earlier prompts.

\subsubsection{Clarification Following AI-Induced Misunderstanding}

We define a second behavioural indicator of cognitive difficulty capturing explicit clarification requests made by users during the conversation. We distinguish between clarification arising from difficulty with the underlying task or subject matter and clarification following difficulty understanding or making sense of the AI response. This distinction allows us to isolate clarification behaviour more directly associated with the preceding AI response.

We modelled clarification as a multi-label classification task at the user-message level, with two labels:
\begin{itemize}
    \item \emph{Task clarification}: the user encounters difficulty understanding or asks for clarification about the task or subject matter itself, often identifying concepts or definitions that they do not understand (e.g., “what does this term mean”, “help me figure this out”).
    \item \emph{AI-induced misunderstanding}: the user encounters difficulty understanding or interpreting the preceding AI response and explicitly requests clarification or signals that the response is confusing or difficult to make sense of (e.g., “I don’t get your answer”, “this response does not make sense”).

    %\item \emph{Task Clarification} and \emph{AI-induced misunderstanding}: the user
\end{itemize}
Messages could receive both labels simultaneously (e.g., “I still don't understand this term, your answer was confusing”) or none where no clarification occurred (e.g., “compare the two plans”).

Given the scale of the Health and Finance datasets, we automated classification by fine-tuning a DistilBERT model  \cite{sanh2019distilbert} at the user-prompt level. The choice of Transformer-based modelling over rule-based or hard-matching approaches was motivated by the semantic diversity of user prompts, particularly clarification requests, for which contextual representations can capture linguistic variation beyond predefined lexical patterns \cite{devlin-etal-2019-bert}.

As a precursor to fine-tuning, we curated a training dataset of manually annotated user prompts sampled from the broader corpus. To boost representation of the Task clarification and AI-induced misunderstanding labels in the training, we augmented the sample with synthetic examples generated by GPT-5.5 (See Appendix \ref{appendix:training_clarification} for more details on the training dataset). Both synthetic and organic messages were manually reviewed by two co-authors of the paper involved in the labelling process with joint sessions organised to align the application of the classification guidelines.

Once the training dataset was finalised, we trained DistilBERT on 90\% of the data for four epochs (learning rate = 2e-5, batch size = 4) and evaluated it on the remaining 10\%. The best model checkpoint achieved $F1_{Macro}= 0.971$ and this was applied to user messages across the full Health and Finance datasets.

For our analysis, we focus on AI-induced misunderstanding because our primary interest is in difficulty arising from users’ attempts to understand and interpret the AI response, rather than difficulty inherent in the underlying task or subject matter.

% Procedure:
% Create a training dataset: sample user prompts showcasing each behaviour and sample prompts from the datasets absent of this behaviour. If not enough, augment with synthetic samples. Double-check annotations.
% Train a light model in multi-label classification setting: Distilbert
% Apply trained model on full set of user prompts and flag subchats containing at least one user message requesting clarification

\subsection{AI Linguistic Measures}
\label{section:ai_measures}

We developed five linguistic and structural measures characterising properties of AI responses that may contribute to users’ cognitive processing demands: readability, response length, formatting density, lexical diversity, and information density.

We constructed two temporally bounded windows to ensure that the linguistic predictors preceded the corresponding repair event used as a behavioural indicator of cognitive difficulty: (1) a \emph{pre-repetition window} containing all turns before the first user prompt classified as a repetition of an earlier prompt, and (2) a \emph{pre-clarification window} containing all turns before the first user request for clarification following AI-induced misunderstanding.

For the pre-repetition window, the repeated prompt itself and all subsequent turns were excluded. The same temporal convention was applied to the clarification analysis: the event-triggering content and subsequent interaction were excluded from predictor construction.

Unless otherwise stated, metrics were first calculated separately for every eligible turn and then averaged without weighting across turns within the relevant conversation window. This approach gave each conversational turn equal weight rather than weighting conversations toward their longest responses. Analyses that used these measures therefore represent the mean turn-level property of the text observed before an event.

\subsubsection{Readability}
We operationalised readability using Flesch Reading Ease \cite{flesch1948new} \footnote{Flesch Reading Ease was implemented with Python package textstat \url{https://pypi.org/project/textstat/}}, a widely used measure based on word, sentence and syllable count.

% calculated with the `textstat` Python package [version]. 

% For a text containing \(W\) words, \(S\) sentences, and \(Y\) syllables, the measure is:

% \[
% \mathrm{FRE}
% =
% 206.835
% -
% 1.015\left(\frac{W}{S}\right)
% -
% 84.6\left(\frac{Y}{W}\right).
% \]

Higher values indicate easier-to-read text, whereas lower values indicate greater textual difficulty. We calculated Flesch Reading Ease separately for user and assistant turns in each analysis window and averaged valid turn-level scores within a conversation. Empty or missing text was assigned no readability score and was treated as missing in subsequent analyses. We selected this measure because it provides an interpretable indicator of surface-level textual difficulty; it does not directly measure conceptual or clinical complexity.

\subsubsection{AI Response Length}
%We measured response length as the number of tokens matching the regular expression `\w+`, which captures alphanumeric word-like sequences. 
We measured response length as the number of alphanumeric word-like sequences identified through regular expressions. Word counts were calculated separately for each user and assistant turn. When a window contained multiple turns, we used the unweighted mean number of words per turn. Accordingly, this measure represents average turn length rather than the total volume of text in the conversation window.

\subsubsection{Formatting Density}
To characterise the visual organisation of assistant responses, we used regular expressions to detect several types of formatting elements: lists (both numbered and unordered), markdown headings, bold spans of text, formatting emojis (e.g., tick, pointing hand, star).

% - unordered-list items beginning with `-`, `*`, or `•`;
% - numbered-list items;
% - Markdown headings at levels 1–6;
% - spans marked as bold using `**text**` or `\_\_text\_\_`; and
% - the formatting emojis such as tick, pointing hand, star, etc.

For each assistant turn, we summed the detected elements and normalized the result by word count:

\[
D_{\mathrm{format}}
=
\frac{30E}{W},
\]

where \(E\) is the number of detected formatting elements and \(W\) is the number of word tokens. The resulting measure represents formatting elements per 30 words. Turns containing no word tokens were assigned a formatting-density value of zero. Turn-level densities were averaged without weighting within each analysis window.

This operationalization captures explicit Markdown-like structure but does not capture all possible visual organization, such as whitespace, indentation not associated with list syntax, or formatting introduced by the user interface.

\subsubsection{Lexical Diversity}
We measured lexical diversity using the Measure of Textual Lexical Diversity (MTLD) \cite{mccarthy2010mtld}.

Text was tokenized using NLTK’s English word tokenizer \footnote{ \url{https://www.nltk.org/}}. Tokens containing no alphanumeric character were removed. MTLD estimates how many tokens a text maintains before its cumulative type–token ratio falls below a specified threshold. We used the conventional threshold \(\tau=0.72\).

For each directional pass through a turn, we incremented a lexical factor whenever the cumulative type–token ratio fell below 0.72 and then restarted the calculation. For an incomplete final segment with type–token ratio \(TTR_f<1\), the partial factor was:

\[
F_{\mathrm{partial}}
=
\frac{1-TTR_f}{1-0.72}.
\]

We ran the procedure in both the original and reversed token orders. The final score was:

\[
\mathrm{MTLD}
=
\frac{N_{\mathrm{tokens}}}
{\left(F_{\mathrm{forward}}+F_{\mathrm{backward}}\right)/2}.
\]

Higher scores indicate that lexical diversity is sustained across longer sequences. MTLD was calculated for each assistant turn and averaged within each analysis window. Empty inputs were assigned a value of zero. Computed values were rounded to four decimal places before export.

\subsubsection{Information Density}

We estimated information density as the ratio of content words to function words. Tokens were assigned Penn Treebank part-of-speech tags using NLTK’s English perceptron tagger. We classified nouns, verbs, and adjectives as content words, while others were classified into function-word categories.

% beginning with `NN`, `VB`, or `JJ`—as content words. We classified the following tags as function-word categories:

% \[
% \{\texttt{PRP, PRP\$, WP, WP\$, IN, CC, DT, EX, TO, MD, POS, RP, WDT, PDT}\}.
% \]

For each assistant turn, information density was calculated as:

\[
D_{\mathrm{information}}
=
\frac{N_{\mathrm{content}}}
{N_{\mathrm{function}}}.
\]

If the part-of-speech tagger identified no function words, we used membership in NLTK’s English stop-word list as a fallback estimate of the denominator. The metric was recorded as missing if the function-word count remained zero. Valid turn-level ratios were averaged within the conversation window and rounded to four decimal places.

This measure should be interpreted as a computational proxy for lexical information density, rather than a direct measure of the factual accuracy or clinical informativeness of a response.

\subsection{Statistical Modelling}
\label{sec:methods_stat_modelling}
We used population-averaged logistic regression models estimated through generalized estimating equations (GEE) to examine two binary conversational repair outcomes used as behavioural indicators of cognitive difficulty: (1) whether a user produced at least one fragmented repeated prompt and (2) whether a clarification event was classified as following AI-induced misunderstanding.

Because individual users could contribute multiple conversations, models clustered observations by user and assumed an exchangeable within-user correlation structure. We fitted the models using a binomial distribution and logit link in Statsmodels.

%The analysis for both finance and health were restricted to US conversations, with health data containing at least one general health category label. 
 %To reduce sparse intent categories, we combined Automation/Monitoring with Task Execution, Summary/Synthesis with Complex Research, and Transformation with Creation.

We estimated 12 models in total. For Finance, we estimated four models: a main-effects model and an interaction model for each of the two conversational outcomes. For Health, we estimated eight models, repeating the same four specifications using either broad health categories or disease categories as controls.

To examine how AI response characteristics were associated with behavioural indications of cognitive difficulty, we included AI response length, AI-response readability, Markdown-formatting density, lexical diversity measured using MTLD, and information density in all models as predictor variables. We also included user-prompt length and readability as control variables. To account for observed variation in the substantive and task context that may contribute to cognitive demands, the models additionally controlled for Finance or Health categories and for whether conversations covered multiple categories. All models also controlled for user intent and model provider (i.e. ChatGPT or Gemini). LLM provider was represented as a dummy variable, while the non-mutually exclusive Finance, Health, and intent labels were multi-hot encoded. Predictor values were calculated only from interactions preceding the outcome: the repetition model used interactions before the first repeated prompt, whereas the misunderstanding model used interactions before the clarification event. This temporal restriction prevented the event and subsequent conversation from contributing to its predictors. For conversations in which the relevant repair event did not occur, predictors were calculated over the full eligible conversation.

To examine whether relationships between AI-response characteristics and behavioural indications of cognitive difficulty varied with response length, we included interaction terms between AI response length and the other AI-response characteristics, i.e., AI readability, Markdown density, MTLD, and information density. We additionally tested whether the association between AI-response readability and behavioural indicators of cognitive difficulty varied with the readability of the user’s prompt. Interaction effects were visualised as model-predicted probabilities across the observed predictor range, with the moderator fixed at one standard deviation below and above its mean and the remaining numeric covariates fixed at their sample means.

Before fitting the models, Flesch Reading Ease values were constrained to the interval \([-100,120]\), and conversations with fewer than one user-prompt word in either event window were excluded. The notebook computed multivariate extremity using squared Mahalanobis distance over the continuous linguistic measures and removed observations with a distance greater than 500. We report odds ratios and two-sided \(p\)-values.

\section{Results}
%\colorbox{yellow}{Need to check regression coefficients}

We first examine how conversational breakdown differs across user intents, before turning to the characteristics of AI responses associated with repeated prompting and clarification.

\begin{table}[!h]
    \centering
    \tiny
    \caption{GEE logistic regression models for the \emph{Finance} dataset predicting two binary conversational-repair outcomes: (1) fragmented repeated prompting and (2) clarification following AI-induced misunderstanding. Models include characteristics of the LLM, user prompts, AI responses, user intent, and conversation content. Values are reported as odds ratios. Statistical significance: \emph{***} if p < 0.001, \emph{**} if p$\in[0.001,0.01)$, \emph{*} if p $\in [0.01,0.05)$,  \emph{.} if p $\in [0.05,0.1)$}
    \begin{tabular}{p{2.5cm}p{2.5cm}p{1.5cm}p{2.5cm}p{0.8cm}p{0.8cm}p{0.8cm}p{0.8cm}}
    \toprule
  
    & & & & \multicolumn{2}{c}{USER REPEATED} & \multicolumn{2}{c}{CLARIFICATION DUE}\\
    & & & & \multicolumn{2}{c}{PROMPTS} & \multicolumn{2}{c}{AI MISUNDERSTANDING}\\
     &Regression model & & & \textbf{\#1} & \textbf{\#2}& \textbf{\#3}& \textbf{\#4}\\
    
    \midrule
    & \textbf{Predictor}	& \textbf{Base}	& \textbf{Comparison}	& \textbf{Main} &	\textbf{With}& \textbf{Main} &	\textbf{With }\\

    &	&&	& \textbf{effects} &	\textbf{interaction}& \textbf{effects} &	\textbf{interaction}\\

    &	&&	& \textbf{only} &	\textbf{terms}& \textbf{only} &	\textbf{terms}\\
    
    \midrule
    & & & Intercept	& 0.20***	& 0.28***	&0.02***&	0.02***\\
    \midrule
    \textbf{LLM model} & LLM model	& Chatgpt &	Gemini	&4.06***	&4.03***	&2.80***	&2.99***\\
    \midrule
    \textbf{User prompt metrics: average before the first repeated prompt from the user within each chat} &average user prompt word count within a chat 	& numeric &	user prompt word count &	1.00&	1.00	&1.00***	&1.00***\\
     \cmidrule{2-8}
    &user prompt Flesch reading ease	&numeric&	user prompt Flesch reading ease	&0.99***	&0.99***	&0.99**	&1.00\\
    \midrule

    \textbf{AI response metrics: average before the first repeated prompt from the user within each chat} & Flesch Reading Ease &	numeric&	Flesch reading ease&	1.00&	1.00&	1.01*&	1.01.\\
    \cmidrule{2-8}
    & Markdown elements per 30 words &numeric&	markdown count per 30 words&	1.10**&	1.02&	0.91&	0.78.\\
     \cmidrule{2-8}
     &average AI response word count within a chat&	numeric&	AI response word count&	1.00***&	0.99***&	1.00&	0.99***\\
     \cmidrule{2-8}
     &MTLD (Measure of Textual Lexical Diversity)&	numeric&	mtld&	0.99***&	0.99***&	1.00&	1.00*\\
     \cmidrule{2-8}
     &Information density ratio&	numeric&	information density ratio&	1.01&	0.98&	1.02&	0.95\\
     \cmidrule{2-8}
     &Whether the effect of AI r. reading ease is moderated by user prompt reading ease&	interaction&	AI response Flesch reading ease x user prompt Flesch reading ease&& 		1.00&&		1.00*\\
     \cmidrule{2-8}
     &Whether the effect of AI's reading ease is moderated by AI response length&	interaction&	AI response Flesch reading ease x AI word count&& 		1.00&&		1.00**\\
     \cmidrule{2-8}
     &Whether the effect of mark. density is moderated by AI response length&	interaction&	markdown count per 30 words x AI word count&&		1.00*&&		1.00\\
     \cmidrule{2-8}
     &Whether the effect of AI info density is moderated by AI response length&	interaction&	AI info density ratio x AI word count&&		1.00&&		1.00***\\
     \cmidrule{2-8}
     &Whether the effect of AI lexic. diversity is moderated by AI response length&	interaction&	MTLD x AI word count&&		1.00***&&		1.00\\
     \midrule
    \textbf{Overall chat metrics} & Financial categories &	numeric&	number of categories a chat belongs to&  	1.07&	1.06&	0.93&	0.94\\
     \cmidrule{3-8}
     &&unlabelled finance chats&	Benefits \& Public Aid&	0.83&	0.83.&	1.23&	1.22\\
     \cmidrule{3-8}
     &&&Budgeting&	0.91&	0.93&	0.90&	0.89\\
     \cmidrule{4-8}
     &&&Business Finance&	1.49***&	1.49***&	1.70*&	1.74*\\
     \cmidrule{4-8}
     &&&Insurance&	1.04&	1.04&	0.93&	0.93\\
     \cmidrule{4-8}
     &&&Investments&	0.99&	1.00&	1.29.&	1.29.\\
     \cmidrule{4-8}
     &&&Finance Infrastructure&	1.61**&	1.63**&1.07&	1.09\\
     \cmidrule{4-8}
     &&&Payments \& Transfers&	1.01&	1.01&	1.36*&	1.33*\\
     \cmidrule{4-8}
     &&&Retail Banking \& Credit&	0.92&	0.93&	0.96&	0.95\\
     \cmidrule{4-8}
     &&&Tax&	1.03&	1.05&	1.12&	1.10\\
     \cmidrule{2-8}
     &Financial intent&	not this intent&	Execution \& Automation&	0.91&	0.85&	0.78&	0.80\\
     \cmidrule{4-8}
     &&&Comparison \& Optimisation &	1.36***&	1.37***&	1.81***&	1.78***\\
     \cmidrule{4-8}
&&&Financial Problem Resolution&	1.05&	1.07&	2.12***&	2.08***\\
\cmidrule{4-8}
&&&Personal Financial Analysis&	1.56***&	1.59***&	1.66**&	1.70***\\
\cmidrule{4-8}
&&&Financial Planning&	1.35**&	1.27*&	1.02&	0.94\\
\cmidrule{4-8}
&&&Complex Research&	2.13***&	2.16***&	1.28*&	1.30*\\
\cmidrule{4-8}
&&&Simple Retrieval&	0.81***&	0.81***&	0.75**&	0.74**\\
\cmidrule{4-8}
&&&Financial Learning \& Education&	1.80***&	1.81***&	2.57***&	2.59***\\
\cmidrule{4-8}
&&&Creation&	1.60***&	1.63***&	1.65***&1.62***\\
\bottomrule
      
    \end{tabular}
    \label{tab:finance_logit_results}
\end{table}

\begin{table}[!h]
    \centering
    \tiny
    \caption{GEE logistic regression models for the \emph{Health} dataset predicting two binary conversational-repair outcomes: (1) fragmented repeated prompting and (2) clarification following AI-induced misunderstanding. Models include characteristics of the LLM, user prompts, AI responses, user intent, and broad Health content categories. Values are reported as odds ratios. Statistical significance: \emph{***} if p < 0.001, \emph{**} if p$\in[0.001,0.01)$, \emph{*} if p $\in [0.01,0.05)$,  \emph{.} if p$\in [0.05,0.1)$}
    \begin{tabular}{p{2.5cm}p{2.5cm}p{1.5cm}p{2.5cm}p{0.8cm}p{0.8cm}p{0.8cm}p{0.8cm}}
    \toprule
    & & & & \multicolumn{2}{c}{USER REPEATED} & \multicolumn{2}{c}{CLARIFICATION DUE}\\
    & & & & \multicolumn{2}{c}{PROMPTS} & \multicolumn{2}{c}{AI MISUNDERSTANDING}\\
    &Regression model & & & \textbf{\#5} & \textbf{\#6}& \textbf{\#7}& \textbf{\#8}\\
    
    \midrule
    & \textbf{Predictor}	& \textbf{Base}	& \textbf{Comparison}	& \textbf{Main} &	\textbf{With}& \textbf{Main} &	\textbf{With }\\

    &	&&	& \textbf{effects} &	\textbf{interaction}& \textbf{effects} &	\textbf{interaction}\\

    &	&&	& \textbf{only} &	\textbf{terms}& \textbf{only} &	\textbf{terms}\\
    
    \midrule
    & & & Intercept	& 0.15***&	0.19***&	0.02***&	0.03***\\
    \midrule
    \textbf{LLM model} & LLM model	& Chatgpt &	Gemini	&3.89***&	3.80***&	2.42***&	2.48***\\
    \midrule
    \textbf{User prompt metrics: average before the first repeated prompt from the user within each chat} &average user prompt word count within a chat 	& numeric &	user prompt word count &	1.00&	1.00&	1.00***&	1.00***\\
     \cmidrule{2-8}
    &user prompt Flesch reading ease	&numeric&	user prompt Flesch reading ease	&0.99***&	0.99***&	0.99**&	1.00\\
    \midrule

    \textbf{AI response metrics: average before the first repeated prompt from the user within each chat} & Flesch Reading Ease &	numeric&	Flesch reading ease&	1.00&	1.00&	1.01**&	1.01.\\
    \cmidrule{2-8}
    & Markdown elements per 30 words &numeric&	markdown count per 30 words&	1.12**&	1.12.&	0.78***&	0.57***\\
     \cmidrule{2-8}
     &average AI response word count within a chat&	numeric&	AI response word count&	1.00***&	1.00***&	1.00**&	1.00.\\
     \cmidrule{2-8}
     &MTLD (Measure of Textual Lexical Diversity)&	numeric&	mtld&	0.99***&	0.99***&	1.00&	0.99***\\
     \cmidrule{2-8}
     &Information density ratio&	numeric&	information density ratio&	0.98&	0.99&	0.96&	1.04\\
     \cmidrule{2-8}
     &Whether the effect of AI response reading ease is moderated by user prompt reading ease&	interaction&	AI response Flesch reading ease x user prompt Flesch reading ease&& 		1.00&&		1.00\\
     \cmidrule{2-8}
     &Whether the effect of AI response reading ease is moderated by AI response length&	interaction&	AI response Flesch reading ease x AI response word count&& 		1.00.&&		1.00\\
     \cmidrule{2-8}
     &Whether the effect of markdown density is moderated by AI response length&	interaction&	markdown count per 30 words x AI response word count&&		1.00&&		1.00*\\
     \cmidrule{2-8}
     &Whether the effect of AI info density is moderated by AI response length&	interaction&	AI info density ratio x AI response word count&&		1.00&&		1.00\\
     \cmidrule{2-8}
     &Whether the effect of AI lexical diversity is moderated by AI response length&	interaction&	MTLD x AI response word count&&		1.00***&&		1.00**\\
     \midrule
     
    \textbf{Overall chat metrics} & Health categories &	has one broad health category&	has more than one broad health category&  	1.14.&	1.16*&	1.31*&	1.32*\\
    
     \cmidrule{3-8}
     &&not this category&	Clinical \& Medical Knowledge&	1.26**&	1.25**&	1.04&	1.05\\
     \cmidrule{3-8}
     &&&Healthcare Logistics \& System Navigation&	1.11&	1.09&	1.36**&	1.36**\\
     \cmidrule{4-8}
     &&&Lifestyle \& Daily Health Management&	1.46***&	1.45***&	1.24*&	1.25*\\
     \cmidrule{4-8}
     &&&Psychological, Identity \& Self-Perception&	1.40***&	1.39***&	1.50***&	1.50***\\
     \cmidrule{4-8}
     &&&Social, Functional \& Environmental Impact&	1.45***&	1.45***&	1.65***&	1.66***\\
     \cmidrule{2-8}
     &Health intent&	not this intent&	Execution \& Automation&	2.40**&	2.17**&	1.59&	1.18\\
     \cmidrule{4-8}
&&&Comparison \& Optimisation &	1.44***&	1.46***&	0.90&	0.93\\
\cmidrule{4-8}
&&&Health Problem Resolution&	1.11.&	1.13*&	1.28*&	1.30**\\
\cmidrule{4-8}
&&&Health Data Analysis&	1.61***&	1.56**&	2.01**&	2.07**\\
\cmidrule{4-8}
&&&Planning&	1.21&	1.21&	0.59.&	0.58.\\
\cmidrule{4-8}
&&&Complex Research&	1.71***&	1.71***&	1.22.&	1.24.\\
\cmidrule{4-8}
&&&Simple Retrieval&	0.85**&	0.84**&	0.65***&	0.65***\\
\cmidrule{4-8}
&&&Health Learning \& Education&	0.99&	0.99&	1.31*&	1.35*\\
\cmidrule{4-8}
&&&Creation&	1.68***&	1.69***&	1.06&	1.06\\
\bottomrule
      
    \end{tabular}
    \label{tab:health_broad_logit_results}
\end{table}

\begin{table}[!h]
    \centering
    \tiny
    \caption{GEE logistic regression models for the Health dataset predicting two binary conversational-repair outcomes: (1) fragmented repeated prompting and (2) clarification following AI-induced misunderstanding. Models include characteristics of the LLM, user prompts, AI responses, user intent, and disease categories. Values are reported as odds ratios. Statistical significance: \emph{***} if p < 0.001, \emph{**} if p$\in[0.001,0.01)$, \emph{*} if p $\in [0.01,0.05)$,  \emph{.} if p$\in [0.05,0.1)$}
    \begin{tabular}{p{2.5cm}p{2.5cm}p{1.5cm}p{2.5cm}p{0.8cm}p{0.8cm}p{0.8cm}p{0.8cm}}
    \toprule
    
    & & & & \multicolumn{2}{c}{USER REPEATED} & \multicolumn{2}{c}{CLARIFICATION DUE}\\
    & & & & \multicolumn{2}{c}{PROMPTS} & \multicolumn{2}{c}{AI MISUNDERSTANDING}\\
    &Regression model & & & \textbf{\#9} & \textbf{\#10}& \textbf{\#11}& \textbf{\#12}\\
    
    \midrule
    & \textbf{Predictor}	& \textbf{Base}	& \textbf{Comparison}	& \textbf{Main} &	\textbf{With}& \textbf{Main} &	\textbf{With }\\

    &	&&	& \textbf{effects} &	\textbf{interaction}& \textbf{effects} &	\textbf{interaction}\\

    &	&&	& \textbf{only} &	\textbf{terms}& \textbf{only} &	\textbf{terms}\\
    
    \midrule
    & & & Intercept	& 0.18***&	0.22***&	0.03***&	0.03***\\
    \midrule
    \textbf{LLM model} & LLM model	& Chatgpt &	Gemini	&3.78***&	3.70***&	2.53***&	2.58***\\
    \midrule
    \textbf{User prompt metrics: average before the first repeated prompt from the user within each chat} &average user prompt word count within a chat 	& numeric &	user prompt word count &	1.00&	1.00&	1.00***&	1.00***\\
     \cmidrule{2-8}
    &user prompt Flesch reading ease	&numeric&	user prompt Flesch reading ease	&0.99***&	0.99***&	0.99***&	1.00\\
    \midrule

    \textbf{AI response metrics: average before the first repeated prompt from the user within each chat} & Flesch Reading Ease &	numeric&	Flesch reading ease&	1.00&	1.00&	1.01**&	1.02*\\
    \cmidrule{2-8}
    & Markdown elements per 30 words &numeric&	markdown count per 30 words&	1.09*&	1.10&	0.76***&	0.56***\\
     \cmidrule{2-8}
     &average AI response word count within a chat&	numeric&	AI response word count&	1.00***&	1.00***&	1.00*&	1.00.\\
     \cmidrule{2-8}
     &MTLD (Measure of Textual Lexical Diversity)&	numeric&	mtld&	0.99***&	0.99***&	1.00&	0.99**\\
     \cmidrule{2-8}
     &Information density ratio&	numeric&	information density ratio&	0.98&	1.00&	0.97&	1.06\\
     \cmidrule{2-8}
     &Whether the effect of AI response reading ease is moderated by user prompt reading ease&	interaction&	AI response Flesch reading ease x user prompt Flesch reading ease&& 		1.00&&		1.00\\
     \cmidrule{2-8}
     &Whether the effect of AI response reading ease is moderated by AI response length&	interaction&	AI response Flesch reading ease x AI response word count&& 		1.00&&		1.00\\
     \cmidrule{2-8}
     &Whether the effect of markdown density is moderated by AI response length&	interaction&	markdown count per 30 words x AI response word count&&		1.00&&		1.00*\\
     \cmidrule{2-8}
     &Whether the effect of AI info density is moderated by AI response length&	interaction&	AI info density ratio x AI response word count&&		1.00&&		1.00\\
     \cmidrule{2-8}
     &Whether the effect of AI lexical diversity is moderated by AI response length&	interaction&	MTLD x AI response word count&&		1.00***&&		1.00**\\
     \midrule
     
    \textbf{Overall chat metrics} & Health categories &	numeric&	number of categories a subchat belongs to  &  	1.05&	1.04&	0.88&	0.89\\
    
     \cmidrule{3-8}
     &&non-disease health chats&	Infectious \& Parasitic Diseases&	1.40***&	1.41***&	1.28&	1.29\\
     \cmidrule{3-8}
     &&&Non-Communicable Systemic \& Organ-Specific Diseases&	1.35***&	1.34***&	1.11&	1.13\\
     \cmidrule{4-8}
     &&&Endocrine, Metabolic \& Immune&	1.22**&	1.23*&	1.43*&	1.44*\\
     \cmidrule{4-8}
     &&&Oncology (Neoplastic Diseases)&	1.23.&	1.24*&	1.17&	1.16\\
     \cmidrule{4-8}
     &&&Mental, Behavioural \& Neurodev.&	1.39***&	1.40***&	1.86***&	1.85***\\
     \cmidrule{4-8}
     &&&Reproductive, Dermatological \& Sensory Conditions&	1.20.&	1.21*&	1.20&	1.22\\
     \cmidrule{4-8}
     &&&Injury, Poisoning \& Ext. Trauma&	1.44***&	1.44***&	1.52*&	1.53*\\

     \cmidrule{2-8}
&Health intent&	not this intent&	Execution \& Automation&	2.45**&	2.23**&	1.64&	1.22\\
\cmidrule{4-8}
&&&Comparison \& Optimisation &	1.52***&	1.55***&	0.89&	0.93\\
\cmidrule{4-8}
&&&Health Problem Resolution&	1.16*&	1.17**&	1.36**&	1.37***\\
\cmidrule{4-8}
&&&Health Data Analysis&	1.67***&	1.62***&	2.01**&	2.06**\\
\cmidrule{4-8}
&&&Planning&	1.35*&	1.36*&	0.62&	0.61\\
\cmidrule{4-8}
&&&Complex Research&	1.72***&	1.72***&	1.28*&	1.29*\\
\cmidrule{4-8}
&&&Simple Retrieval&	0.83***&	0.83***&	0.61***&	0.61***\\
\cmidrule{4-8}
&&&Health Learning \& Education&	0.96&	0.96&	1.29.&	1.32*\\
\cmidrule{4-8}
&&&Creation&	1.69***&	1.71***&	1.07&	1.07\\
\bottomrule
      
    \end{tabular}
    \label{tab:health_disease_logit_results}
\end{table}

\subsection{Conversational breakdown did not follow a simple gradient with decision authority}

Conversational breakdown differed substantially across user intents, but these differences did not map straightforwardly onto the Inform–Shape–Act decision-authority framework underlying the intent taxonomy. Intents involving greater delegation of judgement were sometimes associated with greater breakdown, but this pattern was not monotonic.

In Finance (see Table \ref{tab:finance_logit_results}), several higher-order decision-support intents were associated with greater odds of conversational breakdown.Complex Research was associated with more than twice the odds of fragmented repeated prompting (OR = 2.13 in $R\#$1; OR = 2.16 in $R\#$2), while Personal Financial Analysis (OR = 1.56 in $R\#$1; OR = 1.59 in $R\#$2) and Comparison \& Optimisation (OR = 1.36 in $R\#$1; OR = 1.37 in $R\#$2) were also associated with elevated odds. Personal Financial Analysis (OR = 1.66 in $R\#$3; OR = 1.70 in $R\#$4) and Comparison \& Optimisation (OR = 1.81 in $R\#$3; OR = 1.78 in $R\#$4) similarly predicted greater clarification following misunderstanding. By contrast, Simple Retrieval, a comparatively bounded information-seeking task, was associated with lower odds of repeated prompting (OR = 0.81 in $R\#$1 and $R\#$2) and clarification (OR = 0.75 in $R\#$3; OR = 0.74 in $R\#$4). 

Financial Learning \& Education sits toward the Inform end of the decision-authority framework, yet was associated with elevated odds of repeated prompting (OR = 1.80 in $R\#$1; OR = 1.81 in $R\#$2) and especially clarification following misunderstanding (OR = 2.57 in $R\#$3; OR = 2.59 in $R\#$4). Simple Retrieval, which also involves relatively little delegation of judgement to AI, showed associations in the opposite direction, with lower odds of both repair outcomes. Although these coefficients do not constitute a direct pairwise comparison between the two intents, their differing patterns are informative. Retrieval typically involves obtaining a bounded piece of information, whereas learning requires the user to understand and integrate information that is, by definition of the task, at least partly unfamiliar. Thus, low decision authority need not imply low cognitive demand.

The Health results (see Table \ref{tab:health_broad_logit_results}) showed the same broad distinction between bounded retrieval and more demanding cognitive work. Complex Research was associated with approximately 1.7 times the odds of repeated prompting (OR = 1.71 in $R\#$5 and $R\#$6) and Health Data Analysis with approximately 1.6 times the odds (OR = 1.61 in $R\#$5; OR = 1.56 in $R\#$6) , while Comparison \& Optimisation was also associated with elevated odds (OR = 1.44 in $R\#$5; OR = 1.46 in $R\#$6). Simple Retrieval was again associated with lower odds of repetition (OR = 0.85 in $R\#$5; OR = 0.84 in $R\#$6).

\subsection{The role of substantive domain differed between Finance and Health}

Beyond user intent, we examined whether conversational breakdown varied by substantive topic. In Finance, relatively few verticals independently predicted breakdown after accounting for intent, model provider, and measured user and response characteristics. The clearest exception was Business Finance, which was associated with higher odds of both repeated prompting (OR = 1.49 in $R\#$1; OR = 1.49 in $R\#$2) and clarification following misunderstanding (OR = 1.70 in $R\#$3; OR = 1.74 in $R\#$4). 

Health showed a more systematic relationship between substantive content and breakdown. Disease-focused conversations generally had higher odds of repeated prompting than non-disease health conversations, with elevated odds across multiple disease categories (see $R\#$9 and $R\#$10 in Table \ref{tab:health_disease_logit_results}). At the broader thematic level, clarification following misunderstanding was particularly elevated for Psychological, Identity \& Self-Perception (OR = 1.50 in $R\#$7; OR = 1.50 in $R\#$8) and Social, Functional \& Environmental Impact (OR = 1.65 in $R\#$7; OR = 1.66 in $R\#$8).

The role of substantive content differed between the two domains. User intent was associated with conversational breakdown across both domains, while associations with substantive content were more consistent in Health than in Finance. Elevated difficulty in Health was not confined to disease-focused or technically medical interactions, but was also observed where health intersected with psychological, social, and functional concerns.

\subsection{Individual response characteristics did not map straightforwardly onto conversational difficulty}

The main-effects models provided little evidence for a simple relationship in which ostensibly more complex AI responses were uniformly associated with greater conversational difficulty.

For fragmented repeated prompting in Finance, greater markdown density was associated with higher odds of repetition (OR = 1.10 in $R\#$1). In contrast, greater lexical diversity was associated with lower odds of repetition (OR = 0.99 in $R\#$1 and $R\#$2), as was greater response length (OR = 0.997 in $R\#$1; OR = 0.995 in $R\#$2). AI response readability and information density were not significantly associated with repetition.

The Health models showed a similar pattern. Greater markdown density was associated with increased odds of repeated prompting (OR = 1.12 in $R\#$5 and $R\#$6), whereas greater response length (OR = 0.997 in $R\#$5 and $R\#$6) and lexical diversity (OR = 0.99 in $R\#$5 and $R\#$6) were associated with reduced odds. AI response readability and information density again showed no significant main associations in the broad-category specification.

These findings challenge simple assumptions about what makes an AI response cognitively demanding. Most notably, greater lexical diversity was associated with less subsequent repeated prompting in both domains. Similarly, longer responses were not necessarily associated with greater difficulty, while greater visual structuring was not necessarily associated with less difficulty. These patterns motivated our interaction models examining how response characteristics operate in combination.

\subsection{The relationship between lexical diversity and repeated prompting depended on response length}

\begin{figure}[h!]
\centering
\begin{subfigure}{.5\textwidth}
  \centering
  \includegraphics[width=.95\linewidth]{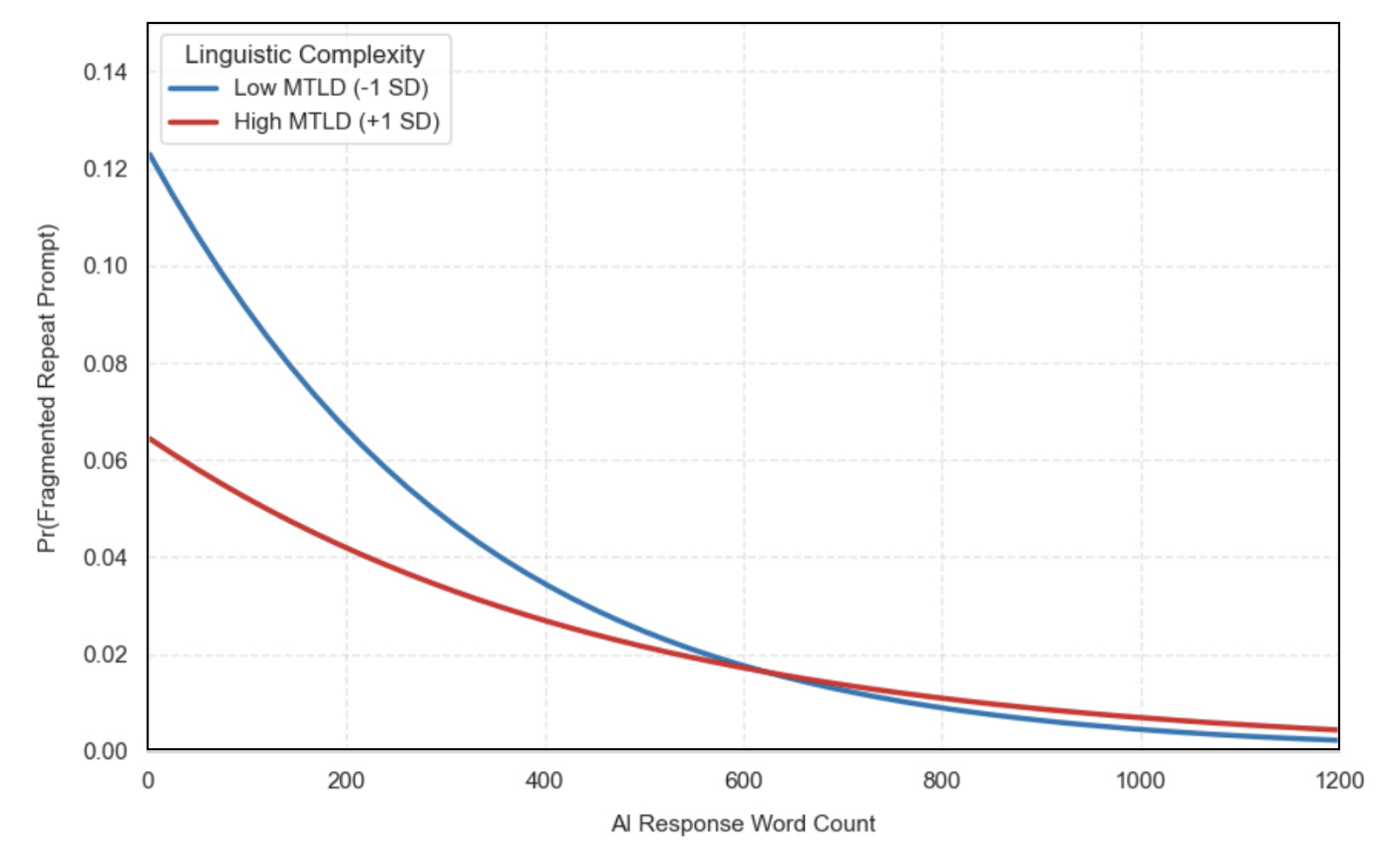}
  \caption{Finance}
  \label{fig:finance_mtldxAI_word_count}
\end{subfigure}%
\begin{subfigure}{.5\textwidth}
  \centering
  \includegraphics[width=.97\linewidth]{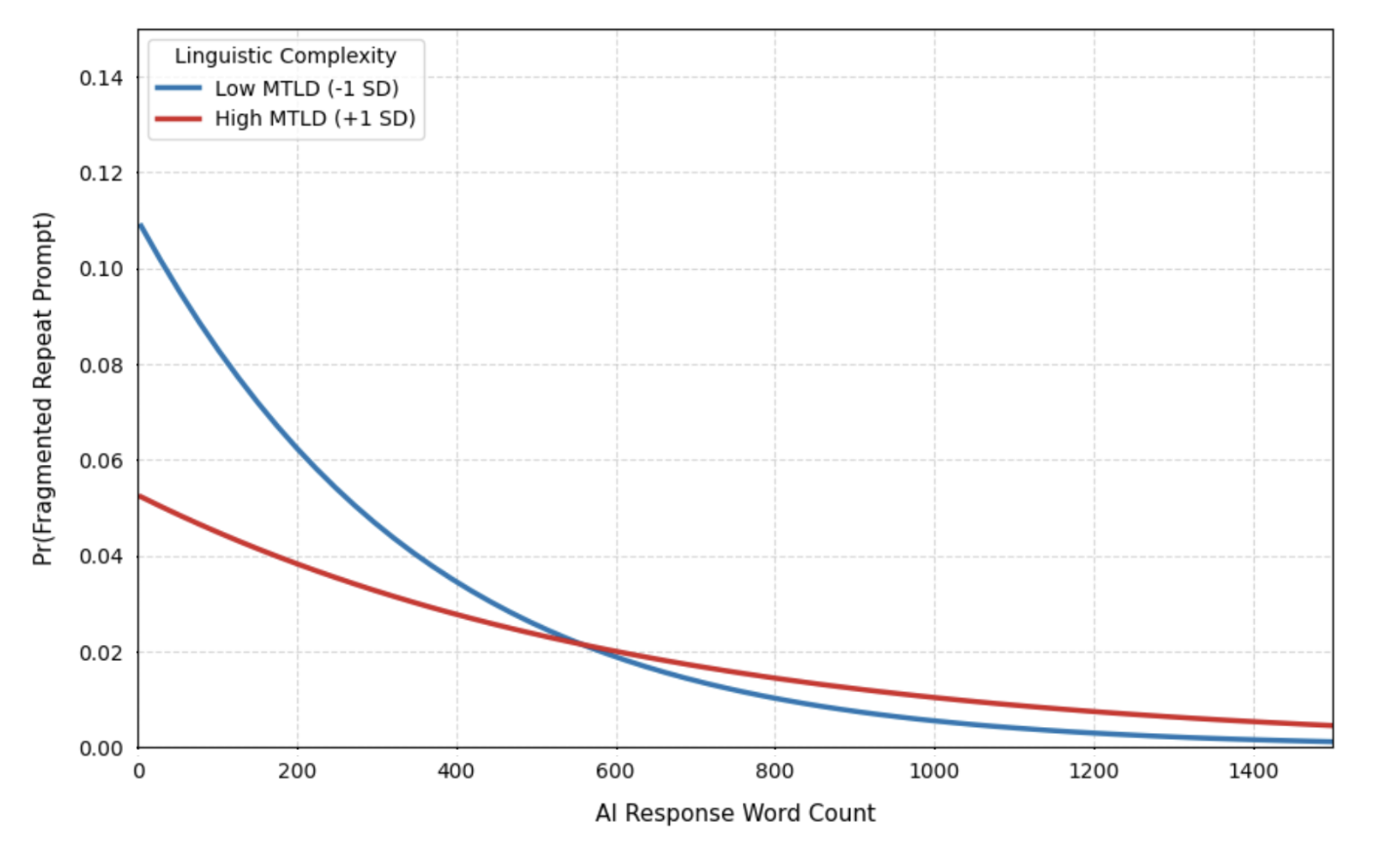}
  \caption{Health}
  \label{fig:health_mtldxAI_word_count}
\end{subfigure}
\caption{Interaction graph between lexical diversity of AI response (signalled by MTLD metric) and its response length (signalled by word count)}
\label{fig:mtldxAI_word_count}
\end{figure}

The clearest cross-domain interaction concerned lexical diversity and response length (see figure \ref{fig:mtldxAI_word_count}).

In Finance, the interaction between MTLD and AI response length significantly predicted fragmented repeated prompting (p < .001; (See $R\#$2 in Table \ref{tab:finance_logit_results})). At shorter response lengths, greater lexical diversity was associated with a substantially lower probability of repeated prompting. This difference progressively narrowed as responses became longer, with the predicted trajectories converging and crossing at approximately 600 words. Beyond this point, the direction of the relationship reversed, although the absolute difference remained comparatively small. This crossover is a model-implied value within the observed data and should not be interpreted as a general threshold for conversational difficulty. 

The MTLD × response-length interaction also significantly predicted repeated prompting in Health (p < .001), and remained significant across alternative specifications controlling for broad health (See $R\#$6 in Table \ref{tab:health_broad_logit_results}) and disease categories (See $R\#$10 in Table \ref{tab:health_disease_logit_results}). The replication of this interaction across domains provides evidence that the association between lexical diversity and conversational difficulty varies with response length.

\subsection{Different configurations of response characteristics predicted explicit misunderstanding}

For clarification in Finance, response length interacted with several other response characteristics.

\begin{figure}[!h]
    \centering
    \includegraphics[width=0.6\linewidth]{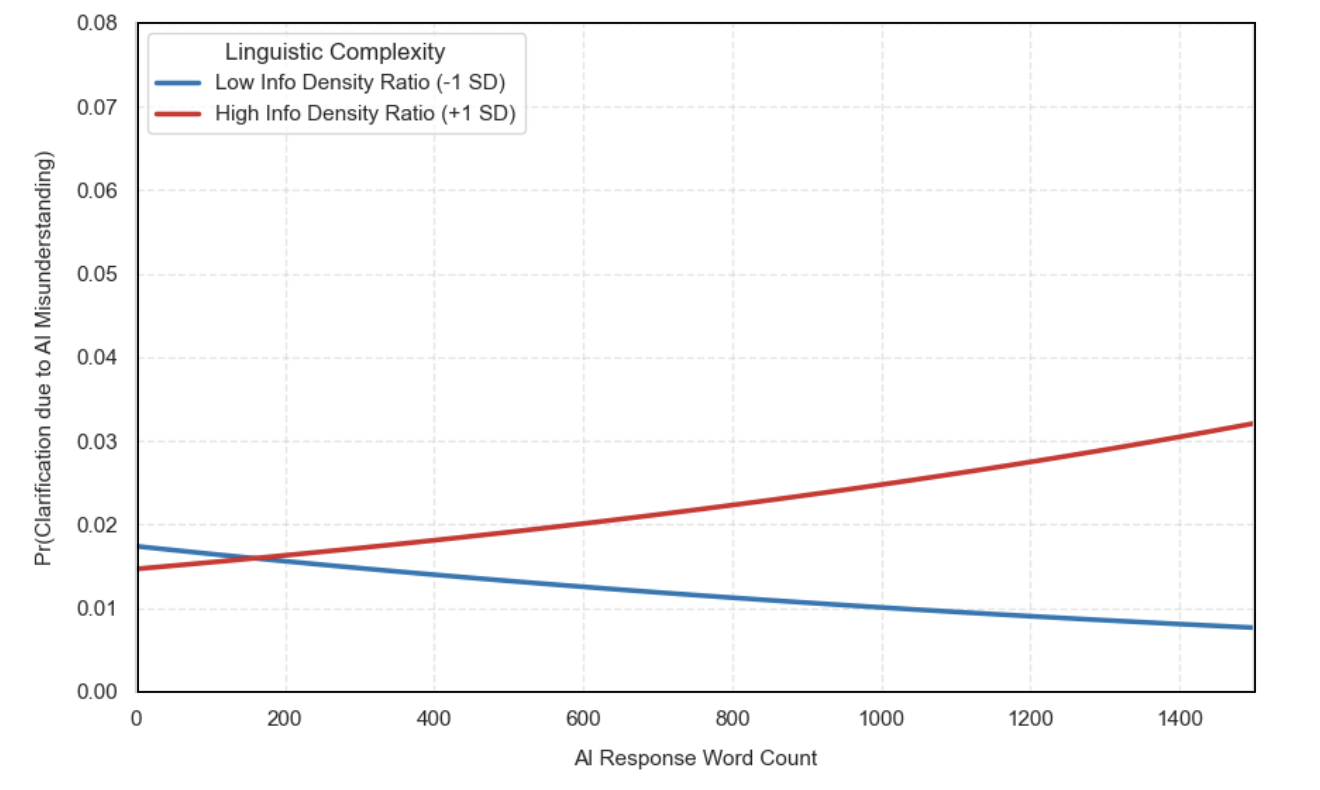}
    \caption{Interaction graph between AI information density and AI response length for Finance conversations}
    \label{fig:finance_info_densityxAI_word_count}
\end{figure}

Information density interacted significantly with response length (See $R\#4$ in Table \ref{tab:finance_logit_results}). At relatively low levels of information density, longer responses were associated with a lower predicted probability of clarification (see Figure \ref{fig:finance_info_densityxAI_word_count}). At high information density, however, this relationship reversed: the predicted probability of clarification increased as responses became longer. The trajectories crossed at approximately 150–200 words. Again, this crossover is specific to the fitted model and observed data and should not be interpreted as a general threshold for conversational difficulty.

\begin{figure}[!h]
    \centering
    \includegraphics[width=0.6\linewidth]{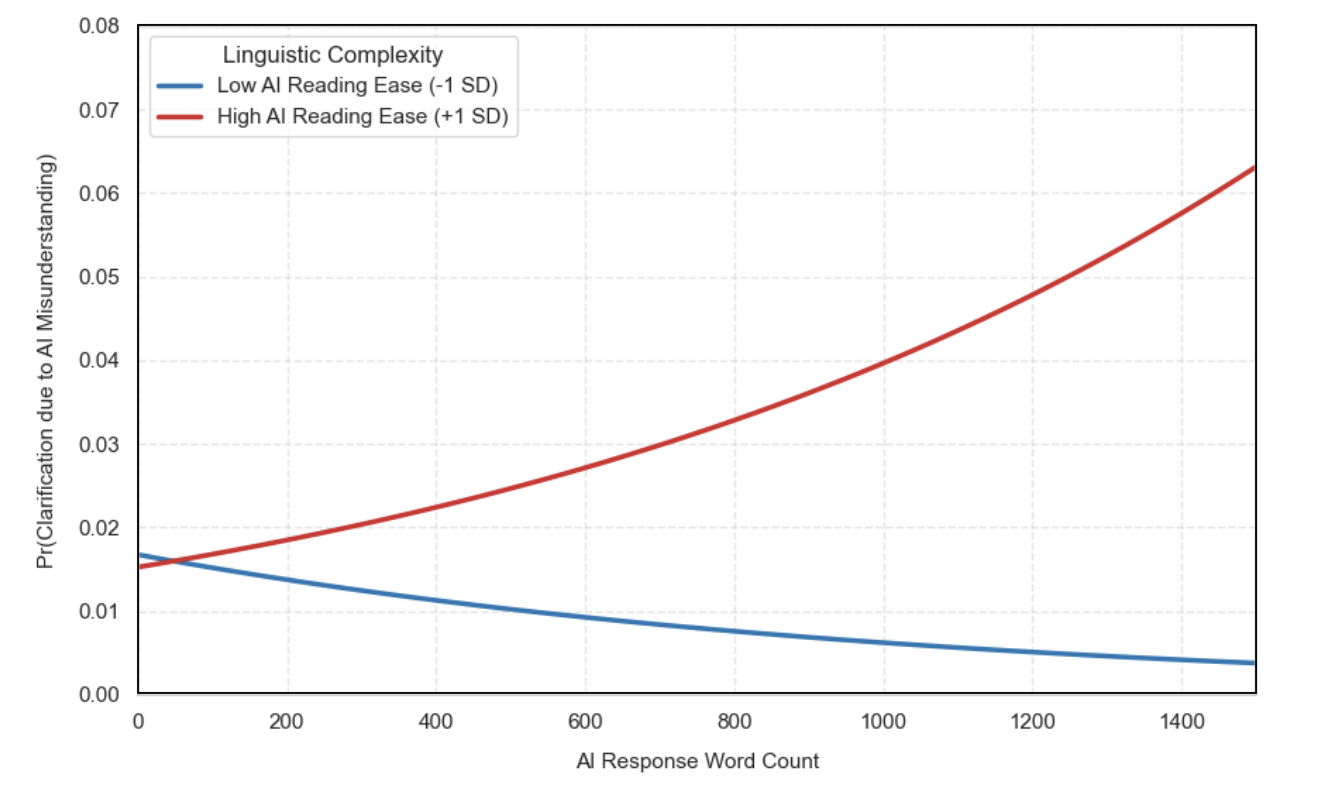}
    \caption{Interaction graph between AI reading ease and AI response length for Finance conversations}
    \label{fig:finance_readabilityxAI_word_count}
\end{figure}

A significant interaction also occurred between AI response readability and response length (See $R\#4$ in Table \ref{tab:finance_logit_results}). As seen in Figure \ref{fig:finance_readabilityxAI_word_count} for responses with relatively low Flesch Reading Ease scores, predicted misunderstanding decreased with response length. Among responses with high reading-ease scores, however, predicted misunderstanding increased as responses became longer, with the difference becoming most pronounced among the longest responses.

These results further demonstrate that response length itself was not consistently associated with greater conversational difficulty. Its relationship with explicit misunderstanding varied with other measured characteristics of the response, including information density and readability.

The readability interaction also highlights a distinction between surface readability and comprehensibility. Flesch Reading Ease captures sentence and word characteristics but does not capture conceptual relationships, qualifications, alternatives or other demands involved in processing a response. One interpretation is that text that is linguistically easy to read may nevertheless impose substantial cognitive demands when presented at greater length.

\begin{figure}[!h]
    \centering
    \includegraphics[width=0.6\linewidth]{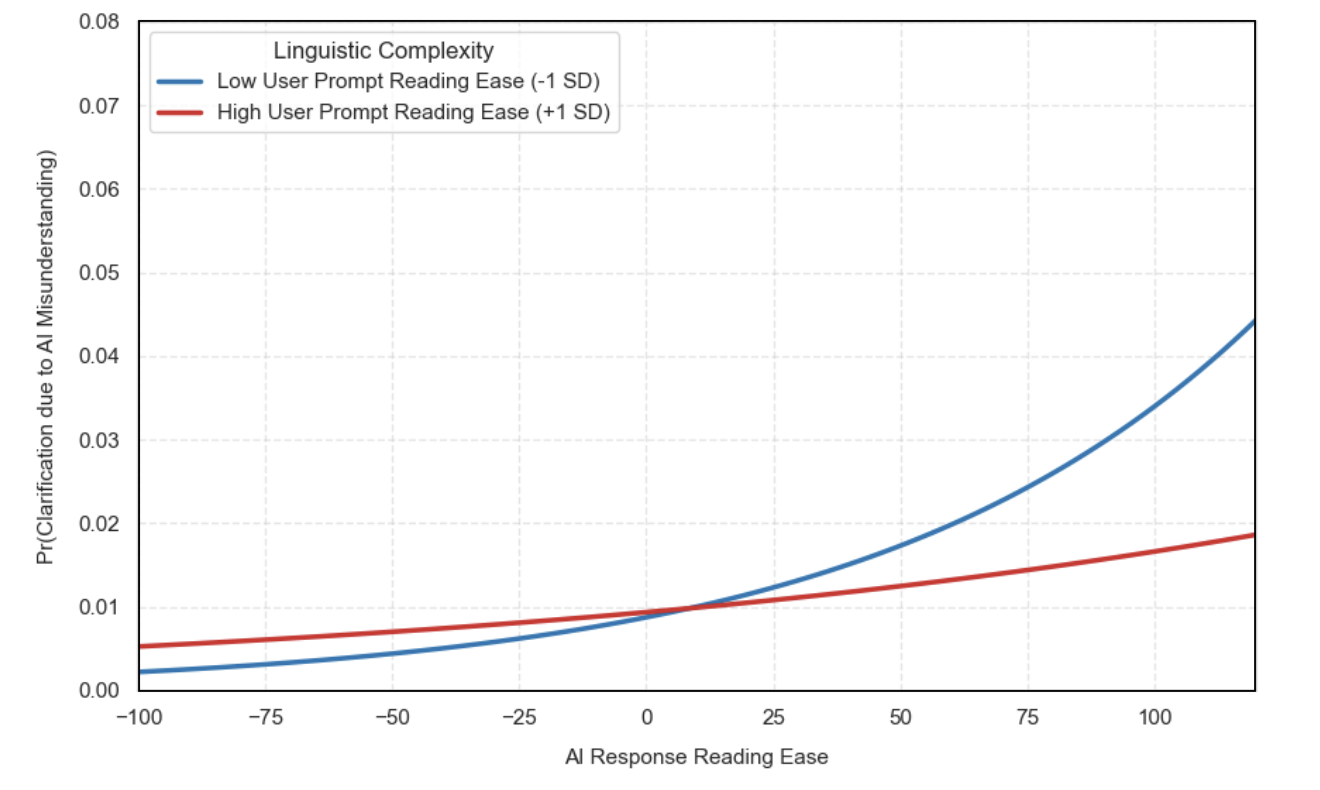}
    \caption{Interaction graph between user prompt's reading ease and AI response's reading ease for Finance conversations}
    \label{fig:finance_readabilityxreadability}
\end{figure}

AI-response readability also interacted with user-prompt readability in Finance (See $R\#$4 in Table \ref{tab:finance_logit_results}). The association between AI reading ease and clarification was stronger when user prompts had relatively low reading-ease scores (see Figure \ref{fig:finance_readabilityxreadability}). We interpret this result cautiously: user-prompt readability should not be treated as a measure of user literacy or ability and may reflect domain terminology, question structure or conversational style. Nevertheless, the result provides preliminary evidence that comprehensibility may depend partly on the relationship between characteristics of the user's input and the system's response, not solely on characteristics of the response itself.

%\colorbox{yellow}{(Insert other relevant interaction graphs for FS and Health)}

\subsection{Model differences persisted after accounting for task and measured response characteristics}

In Finance (see Table \ref{tab:finance_logit_results}), Gemini conversations were associated with approximately four times the odds of fragmented repeated prompting relative to ChatGPT (OR = 4.06 in the main-effects model $R\#$1; OR = 4.03 in the interaction model $R\#$2). For clarification following misunderstanding, the corresponding odds ratios were 2.80 (in the main-effects model $R\#$3) and 2.99 (in the interaction model $R\#$4). The same pattern was observed in Health (see Tables \ref{tab:health_broad_logit_results} and \ref{tab:health_disease_logit_results}). Gemini was associated with approximately 3.7–3.9 times the odds of repeated prompting across specifications (see regressions $\#$ 5, 6, 9 and 10) and approximately 2.5 times the odds of clarification following misunderstanding (see regression $\#$7, 8, 11 and 12).

These differences persisted after accounting for user intent, conversation topic, user-prompt characteristics, and the measured linguistic and structural characteristics of AI responses. Because users were not randomly assigned to models, these estimates should not be interpreted as causal comparisons of model performance.

\section{Discussion}

This study examined when conversational breakdown emerges in real-world human–AI interactions and how it relates to the linguistic and structural characteristics of AI responses. Across more than 84,000 Finance and Health conversations, conversational difficulty could not be reduced to a simple property of either the task or the response. Breakdown differed across user intents, but not simply with the degree of decision authority delegated to AI. Likewise, response characteristics including length, lexical diversity, readability, information density, and formatting did not show uniformly positive or negative relationships with difficulty. Instead, several characteristics interacted, such that their relationship with breakdown depended on other properties of the response.

These findings distinguish decision authority from cognitive demand, suggest that response complexity emerges from configurations of response characteristics, and locate comprehensibility in the broader relationship between user, task, and response.

\subsection{Decision authority and cognitive demand are distinct properties of human–AI interaction}

One way of characterising human–AI interaction is by the degree of decision authority transferred from human to system. The Inform–Shape–Act framework captures this progression, from interactions in which AI supplies information, through those in which it supports analysis and judgement, to those in which it acts with increasing autonomy. Cognitive Load Theory provides a complementary perspective, locating cognitive demand in the processing demands placed on the individual \cite{sweller1988,Sweller2019}. Our findings suggest that these two dimensions do not necessarily move together in human–AI interaction.

Conversational breakdown was generally more common in higher-order interactions involving research, analysis, comparison, and personalised application than during simple retrieval. Financial Learning \& Education provides an important counterexample to a simple decision-authority gradient: despite involving relatively little delegation of judgement, the presence of this intent was associated with elevated breakdown. Together with the broader pattern across intents, this suggests that decision authority delegated to AI and cognitive demands placed on the user are distinct dimensions of human–AI interaction. A user asking AI to explain an unfamiliar concept retains decision authority but must still understand and integrate new information.

\subsection{Defining a conversational complexity budget}

A common intuition in conversational interface design is that cognitive burden can be reduced by making responses shorter, linguistically simpler, and more clearly structured. Our findings complicate this account. More importantly, the interaction models show why individual characteristics provide an incomplete account: the relationship between one response characteristic and breakdown changed depending on what accompanied it.

The clearest example is lexical diversity. Across both Finance and Health, greater lexical diversity was associated with reduced repeated prompting when responses were relatively short, but this advantage diminished as response length increased. Lexical diversity therefore did not have a fixed relationship with difficulty: it was associated with less repeated prompting in shorter responses, but this relationship weakened as response length increased.

The clarification results provide complementary evidence. In Finance, longer responses were associated with lower predicted misunderstanding when information density was low but higher misunderstanding when information density was high. The relationship between surface readability and misunderstanding similarly changed with response length: responses that were easier to read were not necessarily easier to understand when they became long. These findings distinguish the amount of text from the cognitive demands embedded within it. Additional length may provide useful explanatory space under some conditions, while becoming burdensome when combined with other demands.

This pattern is consistent with the concept of element interactivity in Cognitive Load Theory, whereby cognitive demand depends not simply on the amount of information presented, but on the elements that must be processed simultaneously and in relation to one another \cite{Sweller2010,leahy2019centrality}. We extend this logic to the multidimensional structure of conversational AI responses. We propose a conversational complexity budget as a way of conceptualising how demands arising from response characteristics may combine within an interaction. Complexity along one dimension may be readily accommodated when demands along others remain modest; as demands accumulate, the same characteristic may become less helpful or more burdensome. This account is consistent with the replicated lexical-diversity × length interaction and the information-density × length relationship in Finance, but remains a theoretical interpretation; the underlying capacity is not directly measured in our data. Accordingly, we use the term ‘budget’ as a conceptual heuristic for conditional relationships among response demands.

The broader implication is that simplicity should not be treated as a unidimensional design objective. The relevant question may be less whether a response is “simple” or “complex” than whether its different demands can be accommodated together.

\subsection{Evaluating the user–task–response configuration}

So far, we have considered how characteristics within a response combine. But the relevant configuration may extend beyond the response itself: the same response characteristics may relate differently to difficulty depending on the user’s input and task.

The interaction between user-prompt and AI-response readability provides preliminary evidence for this. In Finance, easier-to-read AI responses were associated with more misunderstanding when users' prompts were themselves harder to read, whereas this relationship was much weaker when users' prompts were easier to read. We do not interpret prompt readability as a measure of literacy or ability: it may instead reflect specialist language, question complexity, or conversational style. Nevertheless, the finding suggests that the same style of AI response may work differently depending on the form of the user's input.

This is consistent with longstanding evidence that prior knowledge and expertise shape how people process information \cite{kalyuga2003expertise}, and suggests that similar effects may arise in conversational AI.

This shifts the unit of analysis for conversational AI design. Rather than asking which response characteristics universally minimise cognitive load, HCI research should also ask which response configurations work for which interactions. Comprehensibility, in this account, is relational: it emerges from the fit between the user, the task, and the response.

\subsection{Beyond Universal Simplicity: Designing Adaptive Responses}

These findings suggest a different approach to conversational AI design: adapting response architecture to the demands of the interaction. A general instruction to “be concise” could remove explanatory material that helps users understand unfamiliar information; an instruction to “explain comprehensively” could instead produce responses whose combined length and information density become difficult to process. Prior HCI work has shown that how AI assistance is presented can shape users' cognitive engagement and reliance \cite{bucinca2021}. Our findings extend this work by suggesting that adaptation should target configurations of response demands, not individual response features in isolation. Because our study is observational, these findings do not establish which adaptive response strategies will reduce cognitive difficulty; instead, they motivate a set of testable design hypotheses.

Financial Learning \& Education provides one such case: given its association with elevated breakdown despite low delegation of decision authority, future experiments could compare progressively structured explanations with comprehensive answers delivered at once.

Conversational repair also provides a potential signal for adaptation. Repeated prompting and clarification are observable indications that the current response strategy may not be working. Rather than responding in the same form, a system could adapt by decomposing the task, reducing information density, changing the level of explanation, or asking which part requires clarification.

\subsection{Model differences persisted beyond measured response characteristics}

Substantial differences between models persisted after controlling for user intent, substantive topic, and measured characteristics of user prompts and AI responses. Across both domains, Gemini was associated with approximately four times the odds of repeated prompting and 2.5–3 times the odds of clarification following misunderstanding relative to ChatGPT.

These associations should not be interpreted as causal comparisons of model performance, since users were not randomly assigned to systems. Their persistence indicates that the measured variables in our models do not fully account for the association between model provider and conversational breakdown. The remaining association may reflect unmeasured response properties, such as coherence, explanatory sequencing, contextual alignment, or reasoning structure, as well as differences in users, tasks, interaction histories, interfaces, or model selection.

\subsection{Limitations and future work}

Repeated prompting and clarification are behavioural indicators of cognitive difficulty, not direct measures of cognitive load, and future work should validate them against established cognitive-load metrics. The observational design also precludes causal inference; experiments that independently manipulate response characteristics such as length, information density, and lexical diversity would help establish their causal effects. Finally, our linguistic measures capture only selected dimensions of response complexity, and the analysis is limited to Finance and Health.

Future work should therefore test the conversational complexity budget directly: whether users trade off different forms of response complexity, where breakdown occurs as demands accumulate, and how these relationships vary across users and tasks. This would provide a stronger empirical basis for the adaptive response design proposed here.

\section{Conclusion}

As conversational AI increasingly guides high-stakes activities such as financial and health decision-making, understanding when its responses become difficult to process is an important HCI challenge. Across more than 84,000 real-world conversations, we show that conversational difficulty cannot be explained by individual response characteristics such as length, readability, or lexical diversity alone. Instead, conversational breakdown differs across user intents and is associated with how different characteristics of AI responses combine.

We propose a conversational complexity budget as a framework for understanding these findings: different forms of complexity may be manageable under some conditions, but become more difficult as other demands accumulate. This shifts attention from the response alone toward the broader user–task–response configuration in which complexity is experienced.

For conversational AI design, the challenge is to configure complexity appropriately. Systems that adapt the form and amount of complexity to what users are trying to accomplish, and adjust when conversational breakdown emerges, may better support users in understanding and acting on AI-generated information.

\bibliographystyle{ACM-Reference-Format}
\bibliography{sample-base}

@article{our_paper1,
author = {Bilal, Iman and Wang, Yingcan and Raj, Ajan and Giovagnini, Filippo and Tewari, Pranav and Zhang, Yuwei and Liou, Mei-Chen and Zaman, Qamar},
year = {2026},
month = {08},
pages = {},
title = {From Information to Delegation: Mapping Human-AI Financial Decision Making},
doi = {10.48550/arXiv.2608.02100}
}

@article{chen2025more,
  author = {Chen, Xinyue and Ruan, Kunlin and Ju, Kexin Phyllis and Yap, Nathan and Wang, Xu},
  title = {More AI Assistance Reduces Cognitive Engagement: Examining the AI Assistance Dilemma in AI-Supported Note-Taking},
  journal = {Proceedings of the ACM on Human-Computer Interaction},
  volume = {9},
  number = {CSCW451},
  year = {2025},
  month = {November},
  publisher = {Association for Computing Machinery},
  address = {New York, NY, USA},
  doi = {10.1145/3757632},
  url = {https://doi.org/10.1145/3757632}
}

@article{lepine2026precision,
  title={Precision Proactivity: Measuring Cognitive Load in Real-World AI-Assisted Work}, 
      author={Brandon Lepine and Juho Kim and Pamela Mishkin and Matthew Beane},
      year={2026},
      eprint={2505.10742},
      archivePrefix={arXiv},
      primaryClass={cs.AI},
      url={https://arxiv.org/abs/2505.10742}, 
}

@article{wu2026textual,
  author = {Wu, Siyuan and Peng, Guochao and Li, Shuyang and Cameron, David and Zhang, Jun and Zhang, Zuopeng and Zhang, Qing},
  title = {Textual cues, cognitive load, and social fatigue: Unveiling the reasons behind user discontinuance in conversational AI},
  journal = {International Journal of Human-Computer Studies},
  volume = {215},
  pages = {103846},
  year = {2026},
  month = {June},
  publisher = {Elsevier},
  doi = {10.1016/j.ijhcs.2026.103846},
  url = {https://doi.org/10.1016/j.ijhcs.2026.103846}
}

@article{cruz2026temporal,
  author = {Cruz, Christophe and Jabbar, Samir and Ghanem, Hussam and Bertolim, Maria Alice and Cherifi, Hocine},
  title = {Temporal network analysis of cognitive load and metacognition dynamics in human-AI conversations},
  journal = {Social Network Analysis and Mining},
  year = {2026},
  publisher = {Springer},
  doi = {10.1007/s13278-026-01616-1},
  url = {https://doi.org/10.1007/s13278-026-01616-1}
}

@article{hart1988development,
  author = {Hart, Sandra G. and Staveland, Lowell E.},
  title = {Development of NASA-TLX (Task Load Index): Results of Empirical and Theoretical Research},
  journal = {Advances in Psychology},
  volume = {52},
  pages = {139--183},
  year = {1988},
  doi = {10.1016/S0166-4115(08)62386-9}
}

@article{kalyuga2003expertise,
  author = {Kalyuga, Slava and Ayres, Paul and Chandler, Paul and Sweller, John},
  title = {The expertise reversal effect},
  journal = {Educational Psychologist},
  volume = {38},
  number = {1},
  pages = {23--31},
  year = {2003},
  doi = {10.1207/S15326985EP3801_4}
}

@article{sweller1998cognitive,
  author = {Sweller, John and van Merrienboer, Jeroen J. G. and Paas, Fred G. W. C.},
  title = {Cognitive architecture and instructional design},
  journal = {Educational Psychology Review},
  volume = {10},
  number = {3},
  pages = {251--296},
  year = {1998},
  url = {https://doi.org/10.1023/A:1022193728205}
}

@book{leahy2019centrality,
  author = {Leahy, Wayne and Sweller, John},
  title = {The Centrality of Element Interactivity to Cognitive Load Theory},
  journal = {Advances in Cognitive Load Theory Rethinking Teaching},
  pages = {221--232},
  year = {2019},
  url = {http://dx.doi.org/10.4324/9780429283895-18}
}

@article{hollender2010integrating,
  title = {Integrating cognitive load theory and concepts of human–computer interaction},
journal = {Computers in Human Behavior},
volume = {26},
number = {6},
pages = {1278-1288},
year = {2010},
note = {Online Interactivity: Role of Technology in Behavior Change},
issn = {0747-5632},
doi = {https://doi.org/10.1016/j.chb.2010.05.031},
url = {https://www.sciencedirect.com/science/article/pii/S0747563210001718},
author = {Nina Hollender and Cristian Hofmann and Michael Deneke and Bernhard Schmitz},
}

@inproceedings{bigbird,
author = {Zaheer, Manzil and Guruganesh, Guru and Dubey, Avinava and Ainslie, Joshua and Alberti, Chris and Ontanon, Santiago and Pham, Philip and Ravula, Anirudh and Wang, Qifan and Yang, Li and Ahmed, Amr},
title = {Big bird: transformers for longer sequences},
year = {2020},
isbn = {9781713829546},
publisher = {Curran Associates Inc.},
address = {Red Hook, NY, USA},
articleno = {1450},
numpages = {15},
location = {Vancouver, BC, Canada},
series = {NIPS '20}
}

@article{bucinca2021,
author = {Bu{\c c}inca, Zana and Malaya, Maja Barbara and Gajos, Krzysztof Z.},
title = {To Trust or to Think: Cognitive Forcing Functions Can Reduce Overreliance on AI in AI-assisted Decision-making},
year = {2021},
issue_date = {April 2021},
publisher = {Association for Computing Machinery},
address = {New York, NY, USA},
volume = {5},
number = {CSCW1},
url = {https://doi.org/10.1145/3449287},
doi = {10.1145/3449287},
journal = {Proc. ACM Hum.-Comput. Interact.},
month = apr,
articleno = {188},
numpages = {21}
}

@article{sweller1988,
author = {Sweller, John},
title = {Cognitive Load During Problem Solving: Effects on Learning},
journal = {Cognitive Science},
volume = {12},
number = {2},
pages = {257-285},
doi = {https://doi.org/10.1207/s15516709cog1202\_4},
url ={https://onlinelibrary.wiley.com/doi/abs/10.1207/s15516709cog1202\_4},
eprint = {https://onlinelibrary.wiley.com/doi/pdf/10.1207/s15516709cog1202\_4},
year = {1988}
}

@article{Sweller2019,
  author  = {Sweller, John and van Merri{\"e}nboer, Jeroen J. G. and Paas, Fred},
  title   = {Cognitive Architecture and Instructional Design: 20 Years Later},
  journal = {Educational Psychology Review},
  year    = {2019},
  volume  = {31},
  number  = {2},
  pages   = {261--292},
  doi     = {10.1007/s10648-019-09465-5},
  url     = {https://doi.org/10.1007/s10648-019-09465-5}
}

@article{Sweller2010,
  author  = {Sweller, John},
  title   = {Element Interactivity and Intrinsic, Extraneous, and Germane Cognitive Load},
  journal = {Educational Psychology Review},
  year    = {2010},
  volume  = {22},
  number  = {2},
  pages   = {123--138},
  doi     = {10.1007/s10648-010-9128-5}
}

@article{Schegloffetal1977,
ISSN = {00978507, 15350665},
 URL = {http://www.jstor.org/stable/413107},
 author = {Emanuel A. Schegloff and Gail Jefferson and Harvey Sacks},
 journal = {Language},
 number = {2},
 pages = {361--382},
 publisher = {Linguistic Society of America},
 title = {The Preference for Self-Correction in the Organization of Repair in Conversation},
 urldate = {2026-08-19},
 volume = {53},
 year = {1977}
}

@article{FREEMAN201812,
title = {The effect of response complexity and media on user restatement with multimodal virtual assistants},
journal = {International Journal of Human-Computer Studies},
volume = {119},
pages = {12-27},
year = {2018},
issn = {1071-5819},
doi = {https://doi.org/10.1016/j.ijhcs.2018.06.002},
url = {https://www.sciencedirect.com/science/article/pii/S107158191830315X},
author = {C. Freeman and I. Beaver}
}

@article{DIPPOLD202321,
title = {“Can I have the scan on Tuesday?” User repair in interaction with a task-oriented chatbot and the question of communication skills for AI},
journal = {Journal of Pragmatics},
volume = {204},
pages = {21-32},
year = {2023},
issn = {0378-2166},
doi = {https://doi.org/10.1016/j.pragma.2022.12.004},
url = {https://www.sciencedirect.com/science/article/pii/S0378216622002910},
author = {Doris Dippold}
}

@article{ALLOATTI2024101603,
title = {A tag-based methodology for the detection of user repair strategies in task-oriented conversational agents},
journal = {Computer Speech \& Language},
volume = {86},
pages = {101603},
year = {2024},
issn = {0885-2308},
doi = {https://doi.org/10.1016/j.csl.2023.101603},
url = {https://www.sciencedirect.com/science/article/pii/S0885230823001225},
author = {Francesca Alloatti and Francesca Grasso and Roger Ferrod and Giovanni Siragusa and Luigi {Di Caro} and Federica Cena}
}

@ARTICLE{Parasuraman2000,
  author={Parasuraman, R. and Sheridan, T.B. and Wickens, C.D.},
  journal={IEEE Transactions on Systems, Man, and Cybernetics - Part A: Systems and Humans}, 
  title={A model for types and levels of human interaction with automation}, 
  year={2000},
  volume={30},
  number={3},
  pages={286-297},
  doi={10.1109/3468.844354}}

@misc{friedman2026shortlongllmresponse,
      title={Not Too Short, Not Too Long: How LLM Response Length Shapes People's Critical Thinking in Error Detection}, 
      author={Natalie Friedman and Adelaide Nyanyo and Kevin Weatherwax and Lifei Wang and Chengchao Zhu and Zeshu Zhu and S. Joy Mountford},
      year={2026},
      eprint={2603.06878},
      archivePrefix={arXiv},
      primaryClass={cs.AI},
      url={https://arxiv.org/abs/2603.06878}, 
}

@article{collins2014,
author = {Collins-Thompson, Kevyn},
year = {2014},
month = {12},
pages = {97-135},
title = {Computational assessment of text readability: A survey of current and future research},
volume = {165},
journal = {ITL - International Journal of Applied Linguistics},
doi = {10.1075/itl.165.2.01col}
}

@article{McNamara2012,
place={Cambridge}, 
title={Automated Evaluation of Text and Discourse with Coh-Metrix}, 
publisher={Cambridge University Press}, 
author={McNamara, Danielle S. and Graesser, Arthur C. and McCarthy, Philip M. and Cai, Zhiqiang}, 
year={2014}
}

@inproceedings{sanh2019distilbert,
  title={DistilBERT, a distilled version of BERT: smaller, faster, cheaper and lighter},
  author={Sanh, Victor and Debut, Lysandre and Chaumond, Julien and Wolf, Thomas},
  booktitle={NeurIPS EMC2 Workshop},
  year={2019}
}

@inproceedings{devlin-etal-2019-bert,
    title = "{BERT}: Pre-training of Deep Bidirectional Transformers for Language Understanding",
    author = "Devlin, Jacob  and
      Chang, Ming-Wei  and
      Lee, Kenton  and
      Toutanova, Kristina",
    editor = "Burstein, Jill  and
      Doran, Christy  and
      Solorio, Thamar",
    booktitle = "Proceedings of the 2019 Conference of the North {A}merican Chapter of the Association for Computational Linguistics: Human Language Technologies, Volume 1 (Long and Short Papers)",
    month = jun,
    year = "2019",
    address = "Minneapolis, Minnesota",
    publisher = "Association for Computational Linguistics",
    url = "https://aclanthology.org/N19-1423/",
    doi = "10.18653/v1/N19-1423",
    pages = "4171--4186",
}

@misc{costagomes2025itstimetemporalmodal,
      title={It's About Time: The Temporal and Modal Dynamics of Copilot Usage}, 
      author={Beatriz Costa-Gomes and Sophia Chen and Connie Hsueh and Deborah Morgan and Philipp Schoenegger and Yash Shah and Sam Way and Yuki Zhu and Timothé Adeline and Michael Bhaskar and Mustafa Suleyman and Seth Spielman},
      year={2025},
      eprint={2512.11879},
      archivePrefix={arXiv},
      primaryClass={cs.CY},
      url={https://arxiv.org/abs/2512.11879}, 
}

@article{costagomes2026health,
  author = {Costa-Gomes, Beatriz and Tolmachev, Pavel and Taysom, Eloise and Sounderajah, Viknesh and Richardson, Hannah and Schoenegger, Philipp and Liu, Xiaoxuan and Nour, Matthew M. and Spielman, Seth and Way, Samuel F. and Shah, Yash and Bhaskar, Michael and Nori, Harsha and Kelly, Christopher and Hames, Peter and Gross, Bay and Suleyman, Mustafa and King, Dominic},
  title = {Public use of a generalist LLM chatbot for health queries},
  journal = {Nature Health},
  year = {2026},
  volume = {1},
  number = {7},
  pages = {689--696},
  doi = {10.1038/s44360-026-00117-x},
  url = {https://doi.org/10.1038/s44360-026-00117-x},
  issn = {3005-0693}
}

@article{Beltagy2020Longformer,
  title={Longformer: The Long-Document Transformer},
  author={Iz Beltagy and Matthew E. Peters and Arman Cohan},
  journal={arXiv:2004.05150},
  year={2020},
}

@article{SMAILHODZIC2021114469,
title = {Social media enabled interactions in healthcare: Towards a taxonomy},
journal = {Social Science \& Medicine},
volume = {291},
pages = {114469},
year = {2021},
issn = {0277-9536},
doi = {https://doi.org/10.1016/j.socscimed.2021.114469},
url = {https://www.sciencedirect.com/science/article/pii/S0277953621008017},
author = {Edin Smailhodzic and Albert Boonstra and David J. Langley},
}

@inproceedings{scepanovic2021healthy,
  title={The Healthy States of America: Creating a Health Taxonomy with Social Media},
  author={S{\'c}epanovi{\'c}, Sanja and Aiello, Luca Maria and Zhou, Ke and Joglekar, Sagar and Quercia, Daniele},
  booktitle={Proceedings of the International AAAI Conference on Web and Social Media},
  volume={15},
  pages={690--701},
  year={2021}
}

@Article{rivas2020,
author="Rivas, Ryan
and Sadah, Shouq A
and Guo, Yuhang
and Hristidis, Vagelis",
title="Classification of Health-Related Social Media Posts: Evaluation of Post Content--Classifier Models and Analysis of User Demographics",
journal="JMIR Public Health Surveill",
year="2020",
month="Apr",
day="1",
volume="6",
number="2",
pages="e14952",
issn="2369-2960",
doi="10.2196/14952",
url="https://publichealth.jmir.org/2020/2/e14952",
url="https://doi.org/10.2196/14952",
url="http://www.ncbi.nlm.nih.gov/pubmed/32234706"
}

@article{flesch1948new,
  title={A new readability yardstick},
  author={Flesch, Rudolph},
  journal={Journal of Applied Psychology},
  volume={32},
  number={3},
  pages={221--233},
  year={1948},
  publisher={American Psychological Association},
  doi={10.1037/h0057532}
}

@article{mccarthy2010mtld,
  title={MTLD, vocd-D, and HD-D: A validation study of sophisticated approaches to lexical diversity assessment},
  author={McCarthy, Philip M. and Jarvis, Scott},
  journal={Behavior Research Methods},
  volume={42},
  number={2},
  pages={381--392},
  year={2010},
  publisher={Springer},
  doi={10.3758/BRM.42.2.381}
}

@article{harrison2021icd11,
  author="Harrison, James E.
  and Weber, Stefanie
  and Jakob, Robert
  and Chute, Christopher G.",
  title="ICD-11: an international classification of diseases for the twenty-first century",
  journal="BMC Medical Informatics and Decision Making",
  year="2021",
  month="Nov",
  day="9",
  volume="21",
  number="Suppl 6",
  pages="206",
  issn="1472-6947",
  doi="10.1186/s12911-021-01534-6",
  url="https://doi.org/10.1186/s12911-021-01534-6",
  pubmed="34753471"
}


\appendix
\newpage
\section{Context taxonomy for Finance}
\label{appen:finance_taxonomy}

\begin{table}[h!]
    \centering
    \small
    \caption{Finance taxonomy of Human-AI conversations structured into categories and subcategories by \cite{our_paper1}}
    \label{tab:finance_taxonomy}
    \begin{tabular}{p{4cm}p{5cm}p{4cm}}
    \toprule
    \textbf{Category} & \textbf{Subcategories} & \textbf{Definition}\\
    \midrule
Investments &
\emph{Liquid Securities}
\newline \emph{Alternative Assets}
\newline \emph{Account Types}
\newline \emph{Strategy \& Analysis} & Growth-oriented assets and capital market participation \\
\midrule
 Retail Banking \& Credit &
\emph{Deposit Products}
\newline \emph{Asset-Backed Lending}
\newline \emph{Unsecured Lending}
\newline \emph{Instruments \& Monitoring} & Individual liquidity and consumer debt \\
\midrule

Tax &
\emph{Income \& Output Taxes}
\newline \emph{Asset-Based Taxes}
\newline \emph{Tax Accounting}
 & Statutory obligations and government levies \\
\midrule

Payments \& Transfers &
\emph{Transfer Services}
\newline \emph{Payment Infrastructure}
\newline \emph{Stored Value}
 & The movement of value and transaction processing \\

\midrule
Benefits \& Public Aid &
\emph{Direct Assistance}
\newline \emph{Healthcare Support}
\newline \emph{Disability \& Retirement Support}
\newline \emph{Education Support}
 & Non-market financial support and social safety nets \\

\midrule
Insurance &
\emph{Life \& Health}
\newline \emph{Liability \& Property}
\newline \emph{Policy Mechanics}
 & Contingent contracts for risk transfer \\

  \midrule
Business Finance &
\emph{Commercial Credit}
\newline \emph{Operational Finance}
\newline \emph{Strategic Finance}
 & Corporate and entity-level financial management \\

  \midrule
Finance Infrastructure &
\emph{Regulatory Bodies}
\newline \emph{Verification \& Identity}
\newline \emph{General Infrastructure}
 & Financial ecosystem infrastructure and compliance \\

  \midrule
Budgeting &
\emph{Tracking}
\newline \emph{Planning}
 & Behavioural planning and cash flow management \\
         \bottomrule
    \end{tabular}
\end{table}

\newpage
\section{Context taxonomies for Health}
\label{appen:health_taxonomy}

\begin{table}[h!]
    \centering
    \small
    \caption{General Health taxonomy of Human-AI conversations structured into categories and subcategories inspired by \cite{SMAILHODZIC2021114469,scepanovic2021healthy,rivas2020}}
    \label{tab:general_health_taxonomy}
    \begin{tabular}{p{5cm}p{6cm}p{3cm}}
    \toprule
    \textbf{Category} & \textbf{Subcategories} & \textbf{Definition}\\
    \midrule
Clinical \& Medical Knowledge &
\emph{Symptoms, Manifestations \& Pathophysiology}
\newline \emph{Bio-Medical Interventions}
\newline \emph{Complementary \& Alternative Medicine (CAM)}
\newline \emph{Diagnostics \& Prognosis} & \textit{Focuses strictly on the biological, pharmacological, and physiological aspects of health.} \\
\midrule

Healthcare Logistics \& System Navigation &
\emph{Resource Availability \& Access}
\newline \emph{Institutional Trust \& Patient Advocacy}
\newline \emph{Healthcare Administration \& Policy} & Focuses on the operational, financial, and structural framework of receiving care. \\
\midrule

Lifestyle \& Daily Health Management &
\emph{Nutrition \& Dietary Habits}
\emph{Physical Activity \& Fitness}
\newline \emph{Sleep \& Circadian Hygiene}
\newline \emph{Substance Use \& Harm Reduction}
 & Focuses on non-clinical, day-to-day choices and behaviours that influence well-being. \\
\midrule

Psychological, Identity \& Self-Perception &
\emph{Mental \& Emotional Well-being}
\newline \emph{Body Image \& Physical Self-Perception}
\newline \emph{Health Identity \& Existential Shifts}
 & Focuses on the subjective, internal, and emotional relationship the user has with themselves. \\

\midrule
Social, Functional \& Environmental Impact &
\emph{Interpersonal Relationships \& Support Systems}
\newline \emph{Occupational \& Financial Functioning}
\newline \emph{Caregiving \& Familial Duties}
\newline \emph{Socio-Cultural \& Environmental Influences}
 & Focuses on how health status intersects with the user’s outer world, roles, and relationships. \\
         
         \bottomrule
    \end{tabular}
\end{table}

\newpage
\begin{table}[h!]
    \centering
    \small
    \caption{Disease-specific Health taxonomy of Human-AI conversations structured into categories and subcategories by \cite{harrison2021icd11}}
    \label{tab:disease_health_taxonomy}
    \begin{tabular}{p{5cm}p{6cm}p{4cm}}
    \toprule
    \textbf{Category} & \textbf{Subcategories} & \textbf{Definition}\\
    \midrule
 Infectious \& Parasitic Diseases &
\emph{Viral Infections}
\newline \emph{Bacterial Infections}
\newline \emph{Fungal \& Parasitic Infections} & \textit{Conditions caused by pathogenic microorganisms entering and multiplying in the body.} \\
\midrule

Non-Communicable Systemic \& Organ-Specific Diseases &
\emph{Cardiovascular \& Circulatory Diseases}
\newline \emph{Respiratory Diseases (Non-Infectious)}
\newline \emph{Gastrointestinal \& Hepatic Diseases} 
\newline \emph{Neurological \& Neurodegenerative Disorders}
\newline \emph{Musculoskeletal \& Connective Tissue Disorders} 
\newline \emph{Renal \& Urological Diseases}
\newline \emph{Oral \& Dental Diseases} 
& Chronic or acute conditions affecting specific physiological systems (non-infectious). \\
\midrule

Endocrine, Metabolic \& Immune Disorders &
\emph{Metabolic \& Endocrine Diseases}
\emph{Autoimmune \& Inflammatory Diseases}
\newline \emph{Allergies \& Atopic Conditions}
 & Diseases rooted in hormone imbalances, metabolic dysfunction, or the immune system attacking the body. \\
\midrule

Oncology (Neoplastic Diseases) &
\emph{Solid Tumors (Organ-Specific Cancers)}
\newline \emph{Hematological Malignancies (Blood Cancers)}
\newline \emph{Benign Neoplasms}
 & Benign and malignant tumors, cancers, and cellular overgrowths. \\

\midrule
Mental, Behavioural \& Neurodevelopmental Disorders &
\emph{Mood \& Anxiety Disorders}
\newline \emph{Neurodevelopmental Disorders}
\newline \emph{Personality \& Psychotic Disorders}
\newline \emph{Eating \& Substance Use Disorders}
 & Conditions affecting cognition, emotion, behavioural regulation, and neurological development. \\

 \midrule
Reproductive, Dermatological \& Sensory Conditions &
\emph{Dermatological Conditions (Non-Infectious)}
\newline \emph{Reproductive \& Gynecological Conditions}
\newline \emph{Ophthalmic \& Otolaryngological (Eye/Ear/Throat) Conditions}
 & Conditions localized to external boundaries, reproductive systems, or specialized senses. \\

  \midrule
Injuries, Poisonings \& External Traumas &
\emph{Acute Musculoskeletal Traumas}
\newline \emph{Environmental \& Chemical Injuries}
 & Acute physical damage caused by external forces (often discussed in urgent care AI contexts). \\
         
         \bottomrule
    \end{tabular}
\end{table}

\newpage
\section{Intent taxonomy for Finance and Health}
\label{appen:intent_taxonomy}

\begin{table}[h!]
\centering
\small
  \caption{Taxonomy of user intents in Human-AI conversations with definition, examples and decision authority level.}
  \label{tab:intent_taxonomy}
  \begin{tabular}{p{4.5cm}p{4cm}p{4cm}p{2.5cm}}
    \toprule
    \textbf{Financial Intent} &\textbf{Health Intent} & \textbf{Definition} & \textbf{Decision Authority}\\
    \midrule
    Execution \& Automation \newline
    (previously titled \newline
    Delegated Financial Decision Execution \newline 
    Instruction-led Financial Execution \newline
    Financial Automation \& Monitoring )
    
    & Execution \& Automation
    & Requesting the LLM to execute a task or setting recurring rules, alerts, or optimisations over time & Act\\

\midrule
    Comparison \& Optimisation\newline
    (previously titled \newline
    Product \& Strategy Optimisation\newline
    Product \& Strategy Comparison) & Comparison \& Optimisation & Optimising or comparing one or several recommendations for user & Shape\\
    
\midrule
    Financial Problem Resolution & Health Problem Resolution & Providing solutions for user's issues & Shape\\
\midrule
    Personal Financial Analysis & Health Data Analysis & Interpreting and calculating data based on user's input & Shape \\

\midrule
    Financial Planning & Planning & Structuring future behaviour with step-by-step planning or timelines & Shape\\
    
\midrule
    Complex Research  & Complex Research & Multi-step research that synthesises information across sources, products, or conditions & Inform\\
\midrule
    Simple Retrieval & Simple Retrieval & Retrieving a specific, well-defined piece of information & Inform\\
\midrule
    Financial Learning \& Education & Health Learning \& Education & Building conceptual understanding of a topic, product, or term & Inform\\
\midrule
    Creation & Creation & Generating content (email, code, images) & N/A\\

    \bottomrule
  \end{tabular}
\end{table}
\normalsize

\section{Training corpus for Clarification classification}
\label{appendix:training_clarification}

\begin{table}[h!]
    \centering
    \begin{tabular}{ccccc}
    \toprule
         Label& Organic (count) & Synthetic (count) & Organic (\%) & Synthetic (\%)\\
         \midrule
         AI-induced misunderstanding & 212 & 383 & 35.6 & 64.4\\
         Task Clarification & 476 & 356 & 57.2 & 42.8 \\
         AI-induced misunderstanding, Task Clarification & 110 & 777 & 12.4 & 87.6 \\
         None & 4248 & 0 & 100 & 0\\ 
    \bottomrule
    \end{tabular}
    \caption{Proportion of organic and synthetic data for each of the labels within the clarification classification}
    \label{tab:ai_misunderstanding_training_data}
\end{table}

\end{document}